\documentclass[aps,prb,preprint,superscriptaddress,amsmath,amssymb,citeautoscript,showkeys,endfloats*]{revtex4-2}

\usepackage{amssymb}
\usepackage{amsmath}
\usepackage{upgreek}
\usepackage{bm}
\usepackage{graphicx}
\usepackage{dcolumn}
\usepackage{booktabs}
\usepackage{multirow}
\usepackage{longtable}
\usepackage{color}
\usepackage{xcolor}
\usepackage{hyperref}
\hypersetup{colorlinks = {true},
    		citecolor  = {blue},
    		urlcolor   = {blue},
    		linkcolor  = {blue}}

\newcommand{\Tc}{\ensuremath{T_{\rm c}}}
\newcommand{\Ton}{\ensuremath{T_{\rm c}^{\rm onset}}}
\newcommand{\Tmin}{\ensuremath{T_3}}

\newcommand{\pc}{\ensuremath{p_{\rm c}}}

\newcommand{\rT}{\ensuremath{\rho(T)}}

\newcommand{\LNO}{La$_{3}$Ni$_{2}$O$_{7}$}
\newcommand{\LSNO}{La$_{2.9}$Sr$_{0.1}$Ni$_{2}$O$_{7}$}
\newcommand{\LNNO}{La$_{2}$NdNi$_{2}$O$_{7}$}
\newcommand{\LNSNO}{La$_{1.9}$NdSr$_{0.1}$Ni$_{2}$O$_{7}$}
\newcommand{\LNSSNO}{La$_{1.8}$NdSr$_{0.2}$Ni$_{2}$O$_{7}$}
\newcommand{\LRSNO}{La$_{2-x}R$Sr$_{x}$Ni$_2$O$_7$}
\newcommand{\LNSxNO}{La$_{2-x}$NdSr$_{x}$Ni$_2$O$_7$}
\newcommand{\LSxNO}{La$_{3-x}$Sr$_{x}$Ni$_2$O$_7$}
\newcommand{\RP}{$R_{n+1}M_{n}$O$_{3n+1}$}
\newcommand{\KCO}{KClO$_4$}
\newcommand{\rtenK}{$\rho_{_{\rm 10K}}$}

\newcommand{\dr}{$d\rho/dT$}
\newcommand{\dI}{$d_{x^2-y^2}$}
\newcommand{\dII}{$d_{3z^2-r^2}$}

\newcommand{\Sup}{Supporting Information (SI)}

\begin{document}

%\preprint{}

\author{M.~Kriener}
\email[corresponding author: ]{markus.kriener@riken.jp}
\affiliation{RIKEN Center for Emergent Matter Science (CEMS), Wako 351-0198, Japan}
\author{C.~Terakura}
\affiliation{RIKEN Center for Emergent Matter Science (CEMS), Wako 351-0198, Japan}
\author{A.~Kikkawa}
\affiliation{RIKEN Center for Emergent Matter Science (CEMS), Wako 351-0198, Japan}
\author{Z.~Liu}
\affiliation{RIKEN Center for Emergent Matter Science (CEMS), Wako 351-0198, Japan}
\author{H.~Murayama}
\affiliation{RIKEN Center for Emergent Matter Science (CEMS), Wako 351-0198, Japan}
\author{M.~Nakajima}
\affiliation{RIKEN Center for Emergent Matter Science (CEMS), Wako 351-0198, Japan}
\author{Y.~Fujishiro}
\affiliation{RIKEN Center for Emergent Matter Science (CEMS), Wako 351-0198, Japan}
\affiliation{RIKEN Pioneering Research Institute, Wako, 351-0198, Japan}
\affiliation{Department of Applied Physics, The University of Tokyo, Tokyo 113-8656, Japan}
\affiliation{Institute of Engineering Innovation, The University of Tokyo, Tokyo 113-8656, Japan}
\author{S.~Sasano}
\affiliation{Institute of Engineering Innovation, The University of Tokyo, Tokyo 113-8656, Japan}
\author{R.~Ishikawa}
\affiliation{Institute of Engineering Innovation, The University of Tokyo, Tokyo 113-8656, Japan}
\author{N.~Shibata}
\affiliation{Institute of Engineering Innovation, The University of Tokyo, Tokyo 113-8656, Japan}
\affiliation{Nanostructures Research Laboratory, Japan Fine Ceramics Center, Nagoya, Aichi 456-8587, Japan}
\author{Y.~Tokura}
\affiliation{RIKEN Center for Emergent Matter Science (CEMS), Wako 351-0198, Japan}
\affiliation{Department of Applied Physics, The University of Tokyo, Tokyo 113-8656, Japan}
\affiliation{Tokyo College, The University of Tokyo, Tokyo 113-8656, Japan}
\author{Y.~Taguchi}
\affiliation{RIKEN Center for Emergent Matter Science (CEMS), Wako 351-0198, Japan}
\affiliation{Baton Zone Program, TRIP Headquarters, RIKEN, Wako 351-0198, Japan}

\date{\today}

\title{Controlling the Band Filling and the Band Width in Nickelate Superconductors}

\begin{abstract}
The new family of superconducting nickelates centered around \LNO\ possesses attractive features, such as the high transition temperature and the presence of an antiferromagnetic ground state at ambient pressure, suggesting an unconventional pairing mechanism. In the nonsuperconducting state, the possibility of different density-wave orders with opposite pressure dependencies is discussed, whose relationships and microscopic origins are largely unknown. However, sample-quality issues, such as impurity-phase formation or oxygen vacancies, impede the progress in the field. Here, we employ high-pressure synthesis and hydrostatic high-pressure transport techniques to investigate bilayer nickelates with controlled band width and filling, and perform a systematic study on their impact on the superconductivity and other characteristic properties. While increasing the tilting of the NiO$_6$ octahedra shifts the superconducting phase to higher pressure, simultaneous hole doping reverts this trend. We also observe up to three distinct anomalies in the nonsuperconducting state which are possibly related to density-wave formation.
\end{abstract}


\maketitle
\newpage

%Introduction
\section{Introduction}
The recent discovery of bulk superconductivity in the bilayer nickelate \LNO\ has established a new research field and stimulated tremendous research activity since its superconducting onset temperature under pressure exceeds the boiling point of liquid nitrogen \cite{hsun2023a,jhou2023a,gwang2024a,yzhang2024a,jli2025a,lliu2025a}. \LNO\ belongs to a class of materials called Ruddlesden-Popper (RP) series, \RP, and is composed of double layers ($n=2$) of NiO$_6$ octahedra. Various transition-metal oxides $M$ with the RP structure have been intensively studied as strongly correlated electron systems, particularly in connection with cuprate superconductors, since the 1980s. RP series are well known to offer an ideal platform to control the band width and the band filling independently without disturbing the $M$\,--\,O network by partially changing the $R$ site with an ion with different ionic radius and different valence state, respectively  \cite{imada98a}.

\begin{figure*}[t]
\centering
\includegraphics[width=0.8\linewidth]{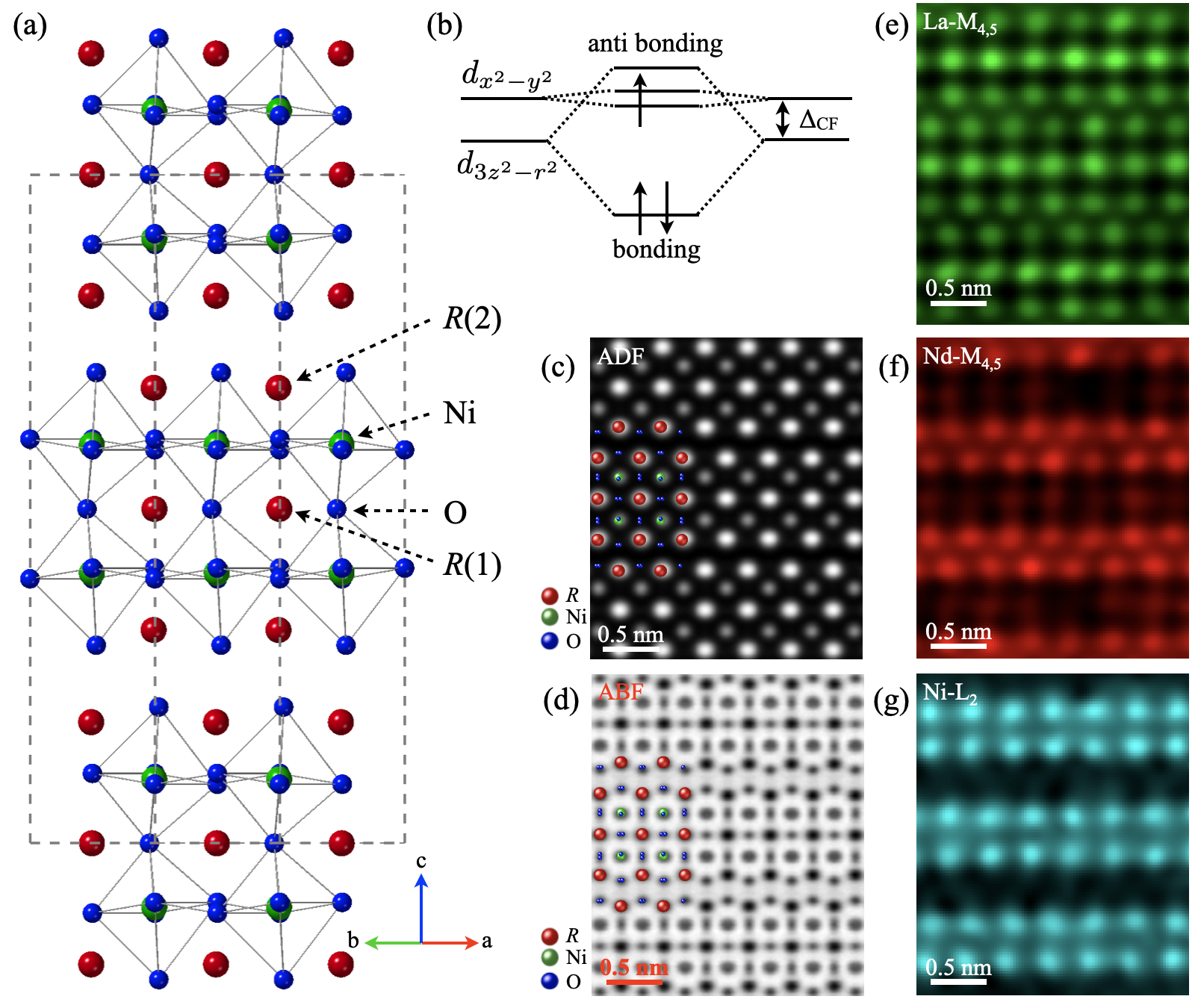}
\caption{\textbf{Crystal and electronic structures.} (a) Ambient-pressure orthorhombic crystal structure with space group $Amam$ of \LNO\ seen along the [110] direction. \LNO\ is characterized by perovskite-like bilayers of tilted NiO$_6$ octahedra, which are separated by rock-salt-like LaO blocks. Crystallographically, there are two distinct La positions labeled $R(1)$ and $R(2)$: $R(1)$ is located within the bilayers between adjacent NiO$_2$ planes and $R(2)$ forms the rock-salt blocks together with the outer apical oxygens. The tilting of the octahedra from the $c$ axis is discernible. The gray dashed lines indicate two unit cells. 
(b) Sketch of the energy levels of the $e_g$ orbitals based on the bilayer model of \LNO\ as proposed in Ref. \cite{sakakibara2024a}. The crystal-field splitting energy $\Delta_{\rm CF}$ denotes the level difference between the \dII\ and \dI\ orbitals. In \LNO, the \dII\ orbitals of two Ni ions in adjacent NiO$_2$ layers hybridize via the inner apical oxygen ions, resulting in bonding-antibonding orbital splitting. 
(c)\,-\,(g) Results of scanning transmission electron microscopy (STEM) on \LNNO: (c) Annular dark- and (d) bright-field STEM images, (e)\,-\,(g) electron energy-loss spectroscopy elemental maps of La, Nd, and Ni, respectively. The atomic structure model is superimposed in panels (c) and (d).} 
\label{fig1}
\end{figure*}

Extensive investigations after the discovery of superconductivity in \LNO\ have revealed that it undergoes structural transitions, as a function of pressure, from the slightly distorted orthorhombic $Amam$ structure at ambient pressure [Fig.~\ref{fig1}(a)] to also orthorhombic $Fmmm$ and eventually to tetragonal $I4/mmm$ structure. The main difference between $Amam$ and $Fmmm$ is that in the former the NiO$_6$ octahedra are tilted from the $c$ axis and are aligned in the latter. This alignment is considered to be a key prerequesite for the emergence of superconductivity \cite{hsun2023a,mwang2024a,lwang2024a,sakakibara2024a}. In the low-pressure region of the phase diagram, \LNO\ exhibits also intriguing features: Charge- and spin-density-wave (CDW and SDW) ordered states with different temperature and pressure dependences are discussed \cite{gwu2001a,zliu2023a,kakoi2024a,kchen2024a,xwyi2024a,ymeng2024a,dzhao2025a,gupta2025a,yashima2025a,khasanov2025a}, although their interplay and microscopic origins remain mostly unresolved.

There are also various theoretical proposals to explain the superconductivity in \LNO\ with the anticipated antiferromagnetic ground state \cite{taniguchi1995a,khasanov2025a}, pointing toward an unconventional pairing mechanism. Currently, many studies suggest favorable $s^{\pm}$-wave pairing, e.g., \cite{ybliu2023a,sakakibara2024a,clu2024a,nomura2025a,ygu2025a,yyzheng2025a}, while others also consider $d$-wave pairing, e.g., \cite{zliao2023a,rjiang2024a,zluo2024a,cxia2025a}. Relevant to the present work is the prediction that partial substitution of La with Nd may enhance \Tc\ \cite{cqchen2026a}. 

Such attempts to partially replace La with divalent Sr or isovalent Pr, Sm, and recently Nd, have been reported. In pristine \LNO, the nominal Ni valence state is $2.5+$, and hence, the electron configuration is $3d^{7.5}$. Figure~\ref{fig1}(b) depicts orbital levels for a single \LNO\ formula unit with large bonding-antibonding splitting of the \dII\ orbitals and small splitting of the \dI\ orbitals while both exhibit a large dispersion when hybridization with the surroundings is further taken into account. Three electrons are accommodated into these four orbitals, and Sr substitution for La dopes a hole into the Ni site. On the other hand, partial substitution of La with smaller and isovalent rare-earth ions increases the lattice distortion, and hence, decreases the bonding-antibonding splitting as well as the hybridization with the surroundings. While only semiconducting behavior is found in La$_{2.8}$Sr$_{0.2}$Ni$_2$O$_7$ \cite{mxu2024a}, the rare-earth-substituted nickelates exhibit superconductivity \cite{nwang2024a,fli2026a,zqiu2025a}. 

Despite these efforts, problems with the reproducibility of experimental results persist, leading to uncertainties and apparent inconsistencies regarding the intrinsic nature of the superconductivity and the nonsuperconducting state from which it emerges. This is reflected, e.g., in wide ranges of the superconducting onset temperature \Ton, the zero-resistance temperature \Tc, and the critical pressure \pc\ required to induce a zero-resistance state \cite{hsun2023a,jhou2023a,yzhang2024a,yzhou2025a,nwang2024a,jli2025a,puphal2024a,fli2026a}, which is mainly due to the formation of different RP impurity nickelate phase intergrowths and oxygen deficiency, e.g., \cite{taniguchi1995a,fli2024a}.

%%%%%%%%%%%%%%%%%%%%%%%%%%%%%%% FIG1 %%%%%%%%%%%%%%%%%%%%%%%%%%%%%%%%%%%
%%%%%%%%%%%%%%%%%%%%%%%%%%%%%% FIG1 %%%%%%%%%%%%%%%%%%%%%%%%%%%%%%%%%%%

Here, we systematically investigate the different effects on the superconducting and nonsuperconducting phases when changing the band width and/or the band filling in \LRSNO\ with $R=$ La ($x = 0$, 0.1) and $R=$ Nd ($x=0$, 0.1, 0.2) grown by a high-pressure synthesis method. The phase diagram of \LNO\ is reproduced, and superconductivity is not observed in hole-doped \LSNO. As for \LNNO\ with an enhanced tilting of the NiO$_6$ octahedra, or a reduced band width, \pc\ is shifted to higher pressure $p$ than in \LNO. Hole doping into \LNNO\ by Sr ($x=0.1$) decreases \pc\ again. Slightly larger $x=0.2$ does not yield bulk superconductivity any more. In the low-pressure region of the phase diagrams, we observe up to three distinct anomalies evolving with $p$ that are possibly related to density-wave instabilities as discussed in the literature.

%%%%%%%%%%%%%%%%%%%%%%%%%%%%%%% FIG2 %%%%%%%%%%%%%%%%%%%%%%%%%%%%%%%%%%%
\begin{figure*}[t]
\centering
\includegraphics[width=1\linewidth]{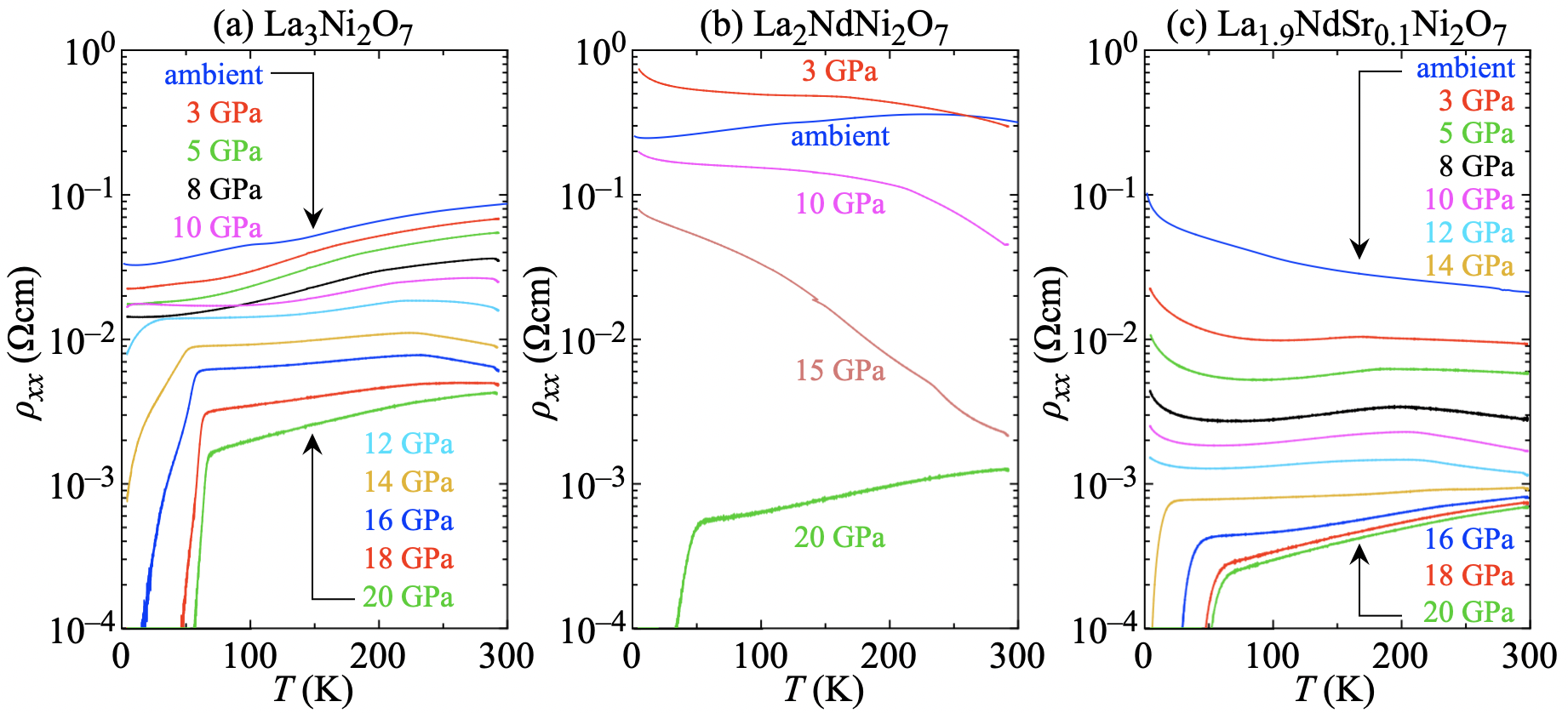}
\caption{\textbf{Transport data of \boldmath{\LRSNO} under pressure.} Temperature-dependent resistivity at various pressures ranging from ambient up to 20~GPa for (a) \LNO, (b) \LNNO, and (c) \LNSNO.} 
\label{fig2}
\end{figure*}
%%%%%%%%%%%%%%%%%%%%%%%%%%%%%% FIG2 %%%%%%%%%%%%%%%%%%%%%%%%%%%%%%%%%%%

%Results
\section{Results}
\label{results}

Figures~\ref{fig1}(c)\,--\,(g) display scanning transmission electron microscopy results on \LNNO, revealing a perfect atomic arrangement of the $n=2$ RP structure without indications of stacking faults or intergrowths of other RP series compositions. Moreover, the intensities of the La and Nd spots in panels (e) and (f) suggest that Nd is preferentially replacing La at the $R(2)$ crystallographic site [shown in Fig.~\ref{fig1}(a)] in the rock-salt blocks between the NiO$_2$ bilayers: The atomic ratios La\,:\,Nd are $(78.8\pm 0.8)$\,:\,$(21.2\pm 0.8)$ and $(61.0\pm 0.6)$\,:\,$(39.0\pm 0.6)$ for the $R(1)$ and $R(2)$ sites, respectively [see Section~S4 in the accompanying \Sup\ \cite{Suppl} for the details and additional images]. This result is qualitatively consistent with the case of $R=\mathrm{Sm}$ \cite{fli2026a}. 

%%%%%%%%%%%%%%%%%%%%%%%%%%%%%%% FIG3 %%%%%%%%%%%%%%%%%%%%%%%%%%%%%%%%%%%
\begin{figure*}[t]
\centering
\includegraphics[width=1\linewidth]{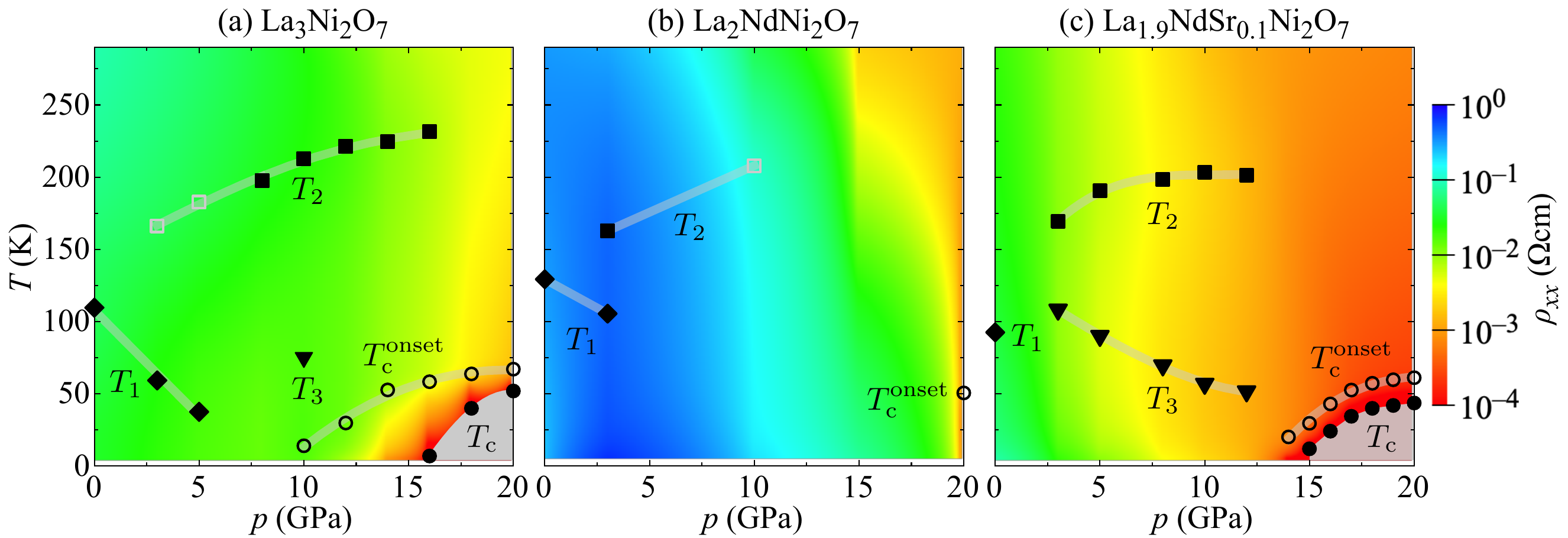}
\caption{\textbf{Temperature-pressure phase diagrams.} Characteristic temperatures \Tc, \Ton, $T_1$, $T_2$, and $T_3$ are plotted in $T-p$ planes for (a) \LNO, (b) \LNNO, and (c) \LNSNO. The superconducting \Tc\ and \Ton\ denote the zero-resistance and onset temperatures upon warming, respectively. In the nonsuperconducting state, $T_1$ indicates anomalies observed in \dr, $T_2$ refers to shoulders or maxima in \rT, and \Tmin\ to  minima in \rT. The $T_2$ anomaly appears with different strength as indicated by filled (clearly discernible) and open symbols (weakly discernible). Thick lines in all panels are guides to the eyes. The background color represents contour plots of the resistivity $\rho(T,p)$.}
\label{fig3}
\end{figure*}
%%%%%%%%%%%%%%%%%%%%%%%%%%%%%% FIG3 %%%%%%%%%%%%%%%%%%%%%%%%%%%%%%%%%%%

Figure~\ref{fig2} summarizes temperature-dependent resistivity \rT\ data from ambient $p$ to 20~GPa of (a) \LNO, (b) \LNNO, and (c) \LNSNO. Enlarged views of the raw data are summarized in Section~S5 in \cite{Suppl}, including respective data for \LSNO\ and \LNSSNO. As shown in Fig.~\ref{fig2}(a), at ambient $p$, \rT\ of \LNO\ exhibits a metallic dependence except toward very low $T$. As a function of $p$, $\rho$ is suppressed and remains metallic up to 8~GPa. At 10~GPa, \rT\ exhibits a semiconducting upturn for $T< 100$~K and intersects with the 8-GPa data, highlighting a major change in the transport behavior at low $T$ (cf.\ also Section~S7 in \cite{Suppl}). There is also a discernible drop in $\rho(T)$ below $\Ton \lesssim 14$~K, indicating the emergence of superconductivity. As $p$ increases, \rT\ is further suppressed, \Ton\ is enhanced, and zero resistance is achieved at $16$~GPa, with \Tc\ reaching $52$~K at 20~GPa. 

In the nonsuperconducting state, three types of anomalies are observed in \rT\ data (see also Supplementary Fig.~S6 in \cite{Suppl}): At ambient $p$, there is a wavy feature around 110~K. At 3 and 5~GPa, this feature is difficult to recognize in \rT, but still discernible in the temperature derivatives \dr\ (not shown). It shifts to lower $T$ with $p$ and is no longer detectable for $p\geq 8$~GPa. A distinct anomaly develops for $p\geq 3$~GPa at temperatures $T> 150$~K. Initially, it appears as a shoulder or slope change in \rT, develops into a maximum for $12~{\rm GPa} \leq p\leq 16$~GPa, and fades out above. Consequently, the metallic $T$ dependence for $T>\Ton$ changes into semiconducting behavior above the maxima. Due to the nonmonotonic evolution of \rT\ with $p$, the 10-GPa data exhibits a minimum as third type of anomaly around $\sim 75$~K.

For \LNNO, the pressure dependence of \rT\ shown in Fig.~\ref{fig2}(b) differs from that of \LNO. At ambient $p$, there is already a broad maximum around $\sim 230$~K with a metallic $T$ dependence below, but the absolute value is large ($>10^{-1}~\Omega$cm). At 3~GPa, \rT\ exhibits a semiconducting $T$ dependence and exceeds the ambient-$p$ data below $\sim 260$~K. Upon further increasing $p$, \rT\ is strongly suppressed but retains a semiconducting $T$ dependence. Metallic conduction and an onset of superconductivity are observed at 20~GPa, although a few $\upmu\Omega$cm remain at 4.2~K in spite of the comparatively sharp drop of the resistivity. 

\LNNO\ exhibits only two notable features in the nonsuperconducting state (see also Fig.~S8 in \cite{Suppl}): At ambient $p$, a wavy anomaly appears at $\sim 130$~K. At 3~GPa, this weakens but is still discernible in \dr\ (not shown). Another anomaly is observed at $T > 150$~K in the intermediate $p$ range, albeit less pronounced than in \LNO.

Figure~\ref{fig2}(c) summarizes respective data on \LNSNO. As in \LNO, the resistivity is suppressed as $p$ increases. However, closer inspection of the data reveals noteworthy differences: At ambient $p$, \LNSNO\ exhibits a semiconducting $T$ dependence in the entire $T$ range. This changes drastically at finite $p$. For $3~{\rm GPa} \leq p \leq 12$~GPa, \rT\ exhibits a semiconductor-like $T$ dependence at low- and high-$T$, and metallic behavior in between, forming pronounced minima and maxima. At $p=14$~GPa, a sharp suppression of the resistivity is observed below $\Ton \approx 20$~K, indicating the emergence of superconductivity. Zero resistance is already achieved at slightly higher $p=15$~GPa. Upon further increasing $p$, \Ton\ and \Tc\ enhance, with \Tc\ reaching 43~K at 20~GPa. The transition width $\Ton-\Tc$ is almost constant as a function of $p$, suggesting a rather homogenous sample quality as well as a highly hydrostatic nature of the pressure used in the resistivity measurements. Moreover, the $T$ dependence of the resistivity for $p\geq 14$~GPa is metallic above \Ton, highlighting a $p$-induced semiconductor-to-metal transition when traversing from 12~GPa to 14~GPa at low $T$. This transition occurs at around $\rho \sim 1$~m$\Omega$cm, close to the Ioffe-Regel limit for the resistivity in this system.

As \LNO, \LNSNO\ exhibits three distinct types of anomalies in the nonsuperconducting state (see also Fig.~S9 in \cite{Suppl}): However, no wavy feature is discernible and an anomaly is barely observed only in \dr\ at $\sim 93$~K in the ambient-$p$ data, but not at finite $p$. The pronounced maxima and minima in the data for $3~{\rm GPa} \leq p \leq 12$~GPa appear at $T>150$~K and $T<110$~K, respectively. While the maxima shift to higher $T$ with $p$, the minima exhibit an opposite $p$ dependence.

%%%%%%%%%%%%%%%%%%%%%%%%%%%%%%% FIG4 %%%%%%%%%%%%%%%%%%%%%%%%%%%%%%%%%%%
\begin{figure*}[t]
\centering
\includegraphics[width=0.5\linewidth]{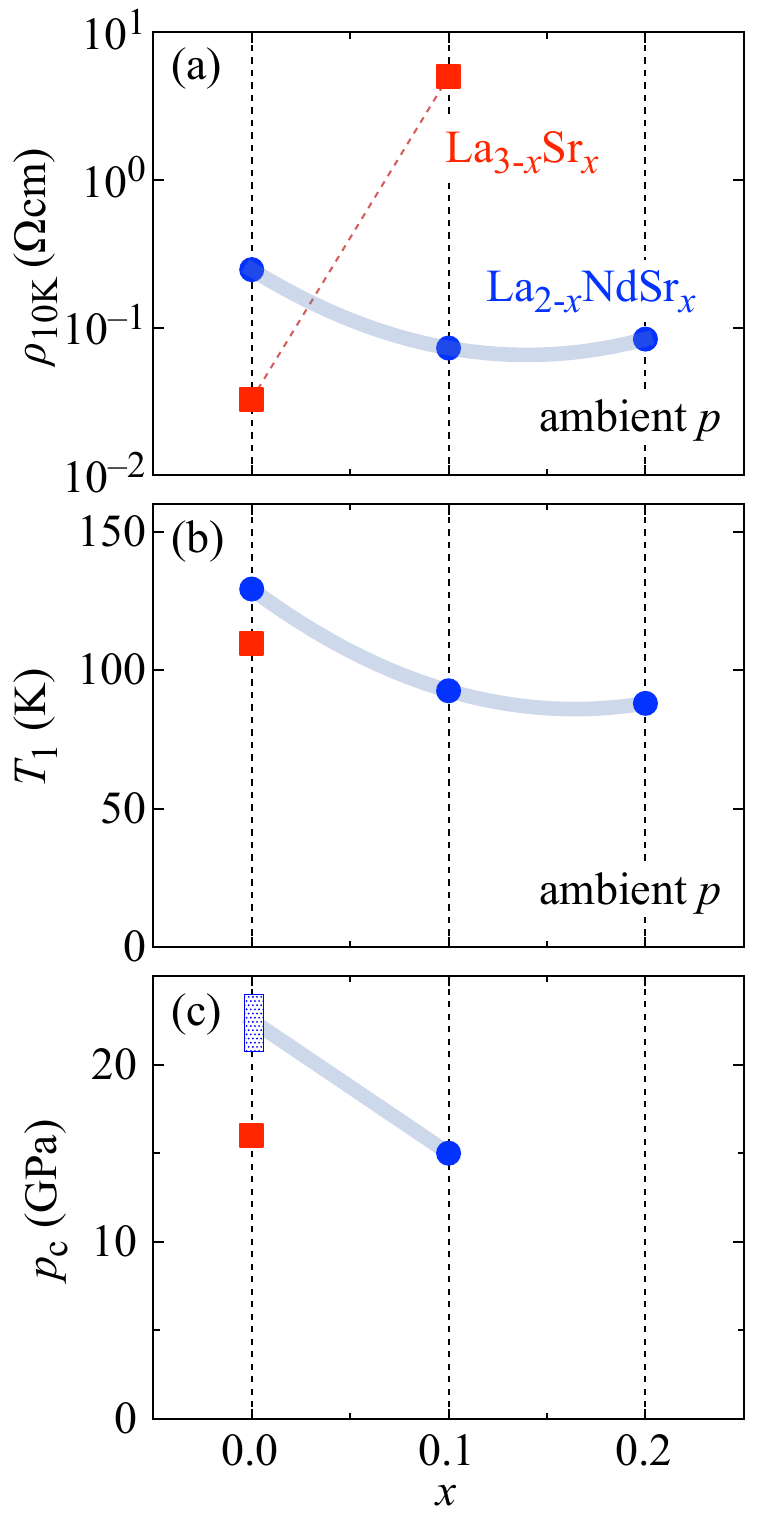}
\caption{\textbf{Characteristic physical quantities of \boldmath{\LRSNO.}} (a) Ambient-pressure resistivity $\rho$ at 10 K, (b) temperature $T_1$ of the anomalies in \dr\ at ambient pressure, and (c) critical pressure \pc\ at which zero resistivity as a function of $p$ is observed, plotted against the Sr concentration $x$. Red and blue symbols refer to \LSxNO\ and \LNSxNO, respectively. The blue dotted rectangular for \LNNO\ in (c) accounts for the observation of a very tiny remaining residual resistivity at 20~GPa. A slightly larger pressure in the $p$ range indicated by the rectangular will likely suppress the resistivity completely. The thin dashed line in (a), the dashed vertical lines highlighting $x$, and the thick lines in all panels are guides to the eyes.}
\label{fig4}
\end{figure*}
%%%%%%%%%%%%%%%%%%%%%%%%%%%%%% FIG4 %%%%%%%%%%%%%%%%%%%%%%%%%%%%%%%%%%%

Figure~\ref{fig3} presents $T-p$ phase diagrams for (a) \LNO, (b) \LNNO, and (c) \LNSNO. The background color in each panel reflects the absolute value of the respective resistivity $\rho(T,p)$. The latter allows an easy assessment of the differences in the evolution leading to the emergence of superconductivity: In \LNO, the resistivity drops to a few m$\Omega$cm (orange-red shading in Fig.~\ref{fig3}) only for $p>\pc$ in a small $T$ range close to \Tc . In \LNNO, the conductivity becomes worse, and the introduction of Nd requires higher pressure to suppress \rT\ and induce superconductivity. By contrast, \LNSNO\ exhibits a sub-m$\Omega$cm resistivity for $p\geq 14$~GPa in the entire $T$ range. 

The $p$ dependences of the characteristic anomalies in the nonsuperconducting state are also plotted in each panel:

(i) All three compositions exhibit an anomaly in \dr\ below 150~K, which is suppressed with increasing $p$ until the first indication of superconductivity is observed. This suggests that its origin competes with the superconductivity. We refer to this type of anomaly as $T_1$ for all three compositions, owing to the similar $T$ range in which it appears and its common negative $p$ dependence. For \LNO, this anomaly has been associated in the literature with a density-wave (DW) instability, sometimes more specifically with a CDW transition \cite{taniguchi1995a,gwu2001a,gwang2024a,khasanov2025a}, suggesting that this is also the case in \LNSxNO.

(ii) Another common feature among the three compositions is observed above 150~K. It manifests itself in \rT\ as shoulders or pronounced maxima in the intermediate $p$ range for \LNO\ and \LNSNO, but appears rather weak in \LNNO. In contrast to the $T_1$ feature, this type of anomaly is enhanced upon increasing pressure until it starts to fade out. For \LNO, it is traceable up to the pressure range where zero resistance is observed, while it vanishes for the two other compositions before the first indication of superconductivity appears. This suggests a competition between its origin and robust superconductivity. Again, on the basis of the similar $T$ range and $p$ dependence, we denote this feature as $T_2$ in all compositions. It remains unclear, however, how the $T_2$ anomalies in Fig.~\ref{fig3} are connected to ambient $p$. According to Ref.~\cite{khasanov2025a}, there is a SDW instability in \LNO\ at ambient $p$ around 150~K which is enhanced as a function of $p$ at least up to $\sim 2.3$~GPa. This anomaly appears to extrapolate into the respective $T_2$ feature shown in Fig~\ref{fig3}(a). Hence, the $T_2$ anomaly observed here may be associated with SDW formation. 

(iii) In the case of \LNO, there is a minimum in $\rho$ as a function of $T$ at 10~GPa. In \LNSNO, a series of pronounced minima is observed in the same pressure range as $T_2$. The difference is that this feature, labeled \Tmin\ in Fig.~\ref{fig3}, shifts to lower $T$ with $p$. However, there is no clear indication that it is related to $T_1$, and the origin of the $T_3$ feature remains unclear at present.

% Discussion
\section{Discussion}
\label{discussion}

Finally, we discuss and compare the results obtained on \LRSNO, on the basis of Fig.~\ref{fig4} which summarizes selected characteristic quantities as a function of the Sr concentration $x$, including results on nonsuperconducting \LSNO\ and \LNSSNO. In Fig.~\ref{fig4}(a), the absolute values of the resistivity \rtenK\ at $T = 10$~K and ambient pressure are shown to quantify the impact of the different compositions on the ground-state conducting properties [A comparative plot of the \rT\ data for all five compositions at ambient $p$ is shown and discussed in Section~S6 in \cite{Suppl}]. \LNO\ exhibits the smallest \rtenK\ in agreement with the metallic \rT\ dependence at ambient $p$. The introduction of smaller isovalent Nd$^{3+}$ is expected to keep the carrier count unchanged, while increasing the lattice distortion and reducing the band width in \LNNO. Hence, \rT\ should increase, as it is observed: \rtenK\ is enhanced by approximately one order of magnitude, and \rT\ increases with $T$ only up to $\sim 230$~K, cf.\ Figs.~\ref{fig2}(a) and (b). 

Introducing divalent Sr$^{2+}$ aims for doping holes and possibly increase the conductivity. In as-grown \LSNO, however, we observe a strongly enhanced \rT\ upon cooling (not shown). This could imply that Sr is not properly incorporated into \LNO, but this is unlikely, given the observed changes in the lattice constants (Fig.~S2 in \cite{Suppl}). Another possibility to explain the large \rT\ is oxygen deficiency. This reduces the Ni valence state toward Ni$^{2+}$, making the system more insulating as experimentally observed in oxygen-deficient \LNO\ \cite{zzhang1994a,taniguchi1995a}. To address this issue, we anneal a sample from the as-grown batch of \LSNO\ in the presence of \KCO, yielding smaller resistivity, but retaining a pronounced semiconducting $T$ dependence, as shown in Fig.~S7 in \cite{Suppl}. The plotted value in Fig.~\ref{fig4}(a) is for this annealed sample. Nevertheless, the resistivity is still very high and superconductivity is not observed up to 20~GPa. This may indicate that either the growth parameters are not yet fully optimized or that the decreased band filling might favor the formation of an insulating state, e.g., due to charge ordering. 

By contrast, substituting Sr$^{2+}$ in \LNSxNO\ has the desired effect and restores conductivity: For $x=0.1$, \rtenK\ decreases. Further increasing the Sr content to $x = 0.2$ yields a slightly larger \rtenK\ than for $x = 0.1$. This might be attributed to a stronger disorder effect, leading to additional localization at low $T$ as shown in Fig.~S11 in \cite{Suppl}. Note that for $x=0.2$, our growth method does not diminish impurity phases as successfully as for $x=0.1$ (Section~S2 in \cite{Suppl}), possibly indicating that $x=0.2$ is close to the solubility limit of Sr in \LNO, even when employing high-pressure synthesis.

Figure~\ref{fig4}(b) summarizes the composition dependence of the $T_1$ anomaly at ambient $p$. For \LNO, we find $T_1 \sim 110$~K. A similar anomaly is observed for \LNNO\ at a higher $T_1 \sim 130$~K, implying that the introduction of smaller Nd$^{3+}$ ions stabilizes the $T_1$ anomaly. If the origin of $T_1$ is related to DW formation, this might be understood as a consequence of enhanced correlation effects, although we cannot rule out that Fermi-surface nesting plays a role, based on the data at hand. By contrast, Sr doping strongly suppresses this feature: The $T_1$ anomaly is absent in \LSNO\ and is shifted to below 100~K in \LNSNO. While one could argue that the large resistivity of \LSNO\ may hide this feature, the comparison with much more conductive \LNSxNO\ suggests that Sr effectively counteracts the origin of the $T_1$ anomaly. It is successively suppressed as a function of $x$ and only barely visible for $x=0.2$ (Section~S6 in \cite{Suppl}) which mimics the physical pressure effect.

Figure~\ref{fig4}(c) presents the composition dependence of the critical pressure \pc\ corresponding to the pressure at which zero resistance is observed. The result for \LNO\ agrees with published data \cite{hsun2023a,gwang2024a,jhou2023a,yzhang2024a}. Whether adding Sr in \LNO\ truly destroys the superconductivity remains an open question, since the large resistivity observed for this batch does not follow a systematic trend, as implied by Figs.~\ref{fig4}(a) and S11 in \cite{Suppl}.

This is further supported by the beneficial effect of adding Sr in \LNNO: While introducing smaller Nd$^{3+}$ ions requires higher pressure to induce zero resistance, hole doping by Sr$^{2+}$ shifts the superconducting phase back to lower pressure. This is in good agreement with the expected opposite effects of Nd$^{3+}$ and Sr$^{2+}$ with modifying the band width and filling, respectively. As can also be seen in Fig.~\ref{fig4}(c), the Sr concentration range in which a robust superconducting phase is established is rather narrow. One possible explanation is that the Sr concentration of 0.2, albeit in favor of reducing the tilting of the NiO$_6$ octahedra, might remove too many electrons from the \dII\ bonding band. According to Ref.~\cite{sakakibara2024a}, too much hole doping of this bonding band suppresses the superconductivity.

As for the anomalies observed at finite pressure, the origin of $T_3$ remains elusive, whereas $T_2$ may be related to a SDW transition, as suggested in the literature \cite{kchen2024a,khasanov2025a}. Interestingly, $T_1$ and $T_2$ exhibit opposite pressure dependences, in contrast to the cuprate case, where DW features are usually coupled. This points to an important difference in the electronic ground states of these families of high-\Tc\ superconductors, i.e., nickelates versus cuprates, and provides an appealing starting point for future experimental and theoretical studies.

\clearpage

\section{Materials and Methods}
\label{methods}

\subsection{Sample growth and characterization}

All batches \LRSNO\ were grown by a high-pressure synthesis method, which requires precursor material. For \LNO, a wet-chemical method was employed, see e.g., \cite{zzhang1994a,taniguchi1995a}, while for the other four compositions, a standard solid-state reaction was used. 

In both cases, the resulting material was thoroughly ground, pressed into pellets, and fired several times under O$_2$ gas flow in a tube furnace to improve the phase purity. 

Ruddlesden-Popper series such as La$_{n+1}$Ni$_{n}$O$_{3n+1}$ are known to suffer from impurity phase formation of other compositions with different $n$ and oxygen vacancies, e.g., \cite{fli2024a,taniguchi1995a}. The final high-pressure synthesis step was implemented here because it has potential to minimize or even remove such impurity phases and, by adding an oxidizing material such as \KCO, to adjust the oxygen stoichiometry. Hence, high-pressure synthesis offers a unique and fast way to approach these issues. 

The growth parameters were optimized for each composition in subsequent high-pressure synthesis attempts and are summarized in Table~S1 in \cite{Suppl}. In the case of \LSNO, the optimized as-grown batch is rather insulating and a post-growth annealing step, also under high pressure and utilizing \KCO, was added to further reduce possible oxygen vacancies. 

All batches were checked with an in-house powder x-ray diffractometer (Rigaku). Respective data are summarized and discussed in Section~S2 in \cite{Suppl}.
The chemical composition and homogeneity of the optimized batches were probed by utilizing a scanning-electron microscope equipped with an energy-dispersive x-ray analyzer (SEM-EDX; JEOL Ltd.\ and Bruker). Selected results are presented in Section~S3 in \cite{Suppl}. 

\subsection{Atomic resolution chemical analysis}
A \LNNO\ sample was gently crushed in ethanol and dispersed on a holey amorphous carbon grid for scanning transmission electron microscopy (STEM) observations. The selected-area electron diffraction was performed using a 2100HC (JEOL Ltd.) operated at 200~kV. Annular dark-field (ADF)- and annular bright-field (ABF) STEM images were acquired using a ARM200F (JEOL Ltd.), equipped with an ASCOR corrector and a cold field-emission gun, operated at 200~kV. The convergence semi-angle was 24~mrad, and the collection semi-angles for ABF and ADF were $12-24$ and $53-170$~mrad, respectively. 

Electron energy-loss spectra (EELS) were acquired by a continuum spectrometer (Gatan Inc.), equipped with an ARM 200F microscope. The convergence and collection semi-angles were 24 and 62~mrad, respectively. To improve the signal-to-noise ratio, a sequential averaging method was used for the ADF-STEM images \cite{ishikawa2014a}, and total third-degree variation (TTDV) regularization was used for EELS elemental maps \cite{kawahara2025a}. In this work, we used the ARIM-mdx data system \cite{hanai2024a} to process the data.

\subsection{Transport measurements}
Rectangular-shaped samples were cut from each batch and polished down to a size of typically $750 \times 550 \times ~450$~\textmu m$^3$ for resistivity measurements under pressure. Four electrical contacts were made with EPO-TEK H20E epoxy. Curing was performed in a two-step process: (i) A sample with fresh contacts was placed on a hot plate ($15\,-\,20$~min at nominally 150$^{\circ}\rm{C}\,-\,200^{\circ}$C in air) to initially cure the EPO-TEK. (ii) It was transferred into a furnace (mini-lamp annealer, Mila 5050Z, Advance Riko) and heated for 1~h at nominally 500$^{\circ}$C in O$_2$ gas flow, cf., e.g., \cite{cyr20218a}. This resulted in two-point resistances of typically a few $\Omega$. Then the sample was measured at ambient pressure in a commercially available system (physical property measurement system, PPMS, Quantum Design) and eventually in a custom-made cubic-anvil press with Daphne oil 7575 as pressure-transmitting medium. Resistivity data under pressure were taken upon warming between $\sim 4.2$~K and $\sim 290$~K. Here, we use a cubic-anvil press to generate a highly hydrostatic pressure distribution.

\section*{Acknowledgement}
We acknowledge the RIKEN TRIP Initiative (Many-Body Electron Systems and Advanced General Intelligence for Science Program). Y.F. acknowledges the support from JST PRESTO (Grant Number JPMJPR2597). A part of this work is supported by “Advanced Research Infra-structure for Materials and Nanotechnology in Japan (ARIM)” of the Ministry of Education, Culture, Sports, Science and Technology (MEXT), (Grant No. JPMXP1225UT0010). S.S. acknowledges the support from JSPS KAKENHI (Grant No. JP 24K17518). R.I. acknowledges the support from JST FOREST (Grant No. JPMJFR2033). N.S. acknowledges the support from JST ERATO (Grant No. JPMJER2202). Fruitful discussions with D.~Maryenko, T.-h.~Arima, X.~Yu, K.~Adachi, and D.~Hashizume are acknowledged.

\section*{Author Contributions}
MK, Y.~Tokura, and Y.~Taguchi conceived the project and designed the experiments. MK, AK, ZL, HM, and MN performed the crystal synthesis. MK characterized the as-grown samples and MK and CT performed the high-pressure transport measurements with YF. MK, Y.~Tokura, and Y.~Taguchi analyzed the data. SS, RI, and NS performed the STEM measurements and analyzed the data. MK wrote the manuscript with contributions from all authors.

\end{document}

%\preprint{}

\author{M.~Kriener}
\email[corresponding author: ]{markus.kriener@riken.jp}
\affiliation{RIKEN Center for Emergent Matter Science (CEMS), Wako 351-0198, Japan}
\author{C.~Terakura}
\affiliation{RIKEN Center for Emergent Matter Science (CEMS), Wako 351-0198, Japan}
\author{A.~Kikkawa}
\affiliation{RIKEN Center for Emergent Matter Science (CEMS), Wako 351-0198, Japan}
\author{Z.~Liu}
\affiliation{RIKEN Center for Emergent Matter Science (CEMS), Wako 351-0198, Japan}
\author{H.~Murayama}
\affiliation{RIKEN Center for Emergent Matter Science (CEMS), Wako 351-0198, Japan}
\author{M.~Nakajima}
\affiliation{RIKEN Center for Emergent Matter Science (CEMS), Wako 351-0198, Japan}
\author{Y.~Fujishiro}
\affiliation{RIKEN Center for Emergent Matter Science (CEMS), Wako 351-0198, Japan}
\affiliation{RIKEN Pioneering Research Institute, Wako, 351-0198, Japan}
\affiliation{Department of Applied Physics, The University of Tokyo, Tokyo 113-8656, Japan}
\affiliation{Institute of Engineering Innovation, The University of Tokyo, Tokyo 113-8656, Japan}
\author{S.~Sasano}
\affiliation{Institute of Engineering Innovation, The University of Tokyo, Tokyo 113-8656, Japan}
\author{R.~Ishikawa}
\affiliation{Institute of Engineering Innovation, The University of Tokyo, Tokyo 113-8656, Japan}
\author{N.~Shibata}
\affiliation{Institute of Engineering Innovation, The University of Tokyo, Tokyo 113-8656, Japan}
\affiliation{Nanostructures Research Laboratory, Japan Fine Ceramics Center, Nagoya, Aichi 456-8587, Japan}
\author{Y.~Tokura}
\affiliation{RIKEN Center for Emergent Matter Science (CEMS), Wako 351-0198, Japan}
\affiliation{Department of Applied Physics, The University of Tokyo, Tokyo 113-8656, Japan}
\affiliation{Tokyo College, University of Tokyo, Tokyo 113-8656, Japan}
\author{Y.~Taguchi}
\affiliation{RIKEN Center for Emergent Matter Science (CEMS), Wako 351-0198, Japan}
\affiliation{Baton Zone Program, TRIP Headquarters, RIKEN, Wako 351-0198, Japan}

\date{\today}

\title{\Sup\ for \\ ``Controlling the Band Filling and the Band Width in Nickelate Superconductors''}

\maketitle

This \Sup\ provides additional data on \LRSNO\ with $R=$\,La ($x = 0$, 0.1) and $R=$\,Nd ($x = 0$, 0.1, 0.2):

\begin{itemize}
\item Section~\ref{growth}: Growth parameters
\item Section~\ref{xrd}: Crystal structure at ambient conditions
\item Section~\ref{edx}: SEM-EDX analyses
\item Section~\ref{stem}: Additional STEM results 
\item Section~\ref{rhocomp}: Expanded plots of $\rho(T,p)$
\item Section~\ref{dramb}: Resistivity $\rho(T)$ and temperature derivatives \dr\ at ambient pressure
\item Section~\ref{nonmon}: Evolution of $\rho(p)$ at $T=10$~K for \LNO\ and \LNSNO
\end{itemize}

\clearpage

\subsection{\textbf{Growth parameters}}
\label{growth}
Table~\ref{tabS1} summarizes the optimized high-pressure synthesis parameters for the five batches discussed in the main text. As mentioned therein, for \LSNO\ an additional post-growth annealing step was carried out for a batch obtained by high-pressure synthesis. Hence, there are two sets of parameters for this composition. 
\begin{table*}[h!]
\label{tabS1}
\centering
\caption{\textbf{High-pressure synthesis parameters for \boldmath{\LRSNO}}}
\setlength{\tabcolsep}{10pt}
\begin{tabular}{cccccc}
\toprule
\multirow{2}*{Composition} & precursor            & $p$   & $T$           & $t$   & \KCO\         \\ \addlinespace[-0.5em]
                           & grown by             & (GPa) & $(^{\circ}$C) & (min) & (molar ratio) \\ \addlinespace[0.5em]\hline \addlinespace[0.5em]
\LNO\                      & wet chemistry        & 3     & 1000          & 60    & --            \\ \addlinespace[0.5em] \hline \addlinespace[0.5em]          % MKR-417HP-1
\multirow{2}*{\LSNO\ }     & i) solid-state reaction & 4     & 1200          & 60    & --            \\ \addlinespace[0.1em]                                      % MKR-421HP-2
                           & ii) high-$p$ synthesis (from i)   & 2     & 650           & 10    & 1/50          \\ \addlinespace[0.5em] \hline \addlinespace[0.5em]          % MKR-421HP-2-OX
\LNNO\                     & solid-state reaction & 5     & 1200          & 60    & --            \\ \addlinespace[0.5em] \hline \addlinespace[0.5em]          % HM-2-HP-7
\LNSNO\                    & solid-state reaction & 5     & 1000          & 60    & 1/40          \\ \addlinespace[0.5em] \hline \addlinespace[0.5em]          % MKR-422HP-8
\LNSSNO\                   & solid-state reaction & 5     & 1000          & 60    & 1/120         \\ \addlinespace[0.5em]                                      % MKR-424HP-7
\bottomrule
\end{tabular}
\end{table*}

\clearpage

\subsection{\textbf{Crystal structure at ambient conditions}}
\label{xrd}
%Figure S1
\begin{figure}[b]
\centering
\includegraphics[width=0.68\linewidth,clip]{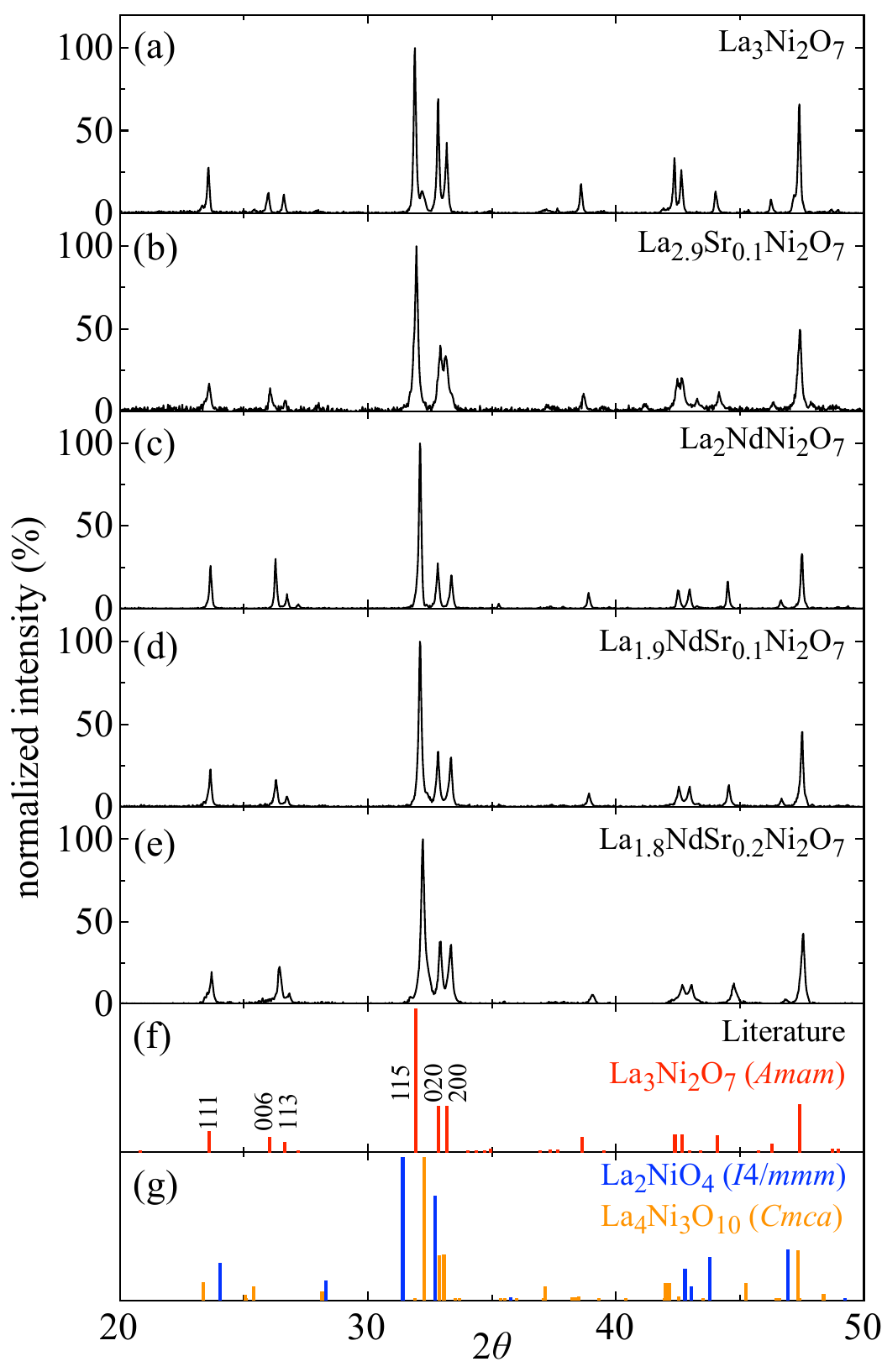}
\caption{(a)\,--\,(e) Normalized x-ray diffraction data of \LRSNO\ with $R =\mathrm{La}$, $x=0$, 0.1, and $R =\mathrm{Nd}$, $x=0$, 0.1, 0.2. Calculated peak positions of (f) \LNO\ (according to Ref.~\cite{ling2000a}, ICSD-91141) and (g) \LNOmo\ (Ref.~\cite{takeda1990a}, ICSD-69172) and \LNOtri\ (Ref.~\cite{voronin2001a}, ICSD-96720).}
\label{figS1}
\end{figure}
Figure~\ref{figS1} summarizes x-ray diffraction (XRD) data at ambient conditions for the five compositions \LRSNO\ along with relevant literature data.

\LNO\ is the bilayer ($n=2$) composition of the Ruddlesden-Popper (RP) series La$_{n+1}$Ni$_{n}$O$_{3n+1}$. RP series are known to suffer from impurity phase formation of related compositions with different $n$. In the present case, phase fractions of the monolayer \LNOmo\ ($n=1$) and the trilayer \LNOtri\ ($n=3$) are often observed, e.g., \cite{nwang2024a,xchen2024a,zdong2024a,fli2024a}. 

After optimizing the high-pressure synthesis parameters, \LNOmo\ is usually not detectable within the resolution limit of the in-house powder x-ray diffractometer used [Figs.~\ref{figS1}(a) -- (d)]. Only in the data for \LNSSNO\ shown in Fig.~\ref{figS1}(e), there is a tiny feature discernible at the left-hand side of the $n=2$ main peak, possibly reflecting a small $n=1$ phase fraction, cf.\ Figs.~\ref{figS1}(e)\,--\,(g). This may indicate that $x=0.2$ is close to the solubility limit of Sr in \LNSxNO. 

In the XRD data on \LNO\ shown in Fig.~\ref{figS1}(a), a small extrinsic peak is seen on the higher-$2\theta$ side of the $n=2$ main peak. It is likely to stem from an \LNOtri\ impurity phase. For the four other compositions, $n=3$ impurity reflections are not clearly discernible due to the shift of the intrinsic peaks induced by chemical substitution. Broadening of the, e.g., $n=2$ main peak in \LNSxNO\  may hint toward the formation of a tiny $n=3$ phase as a function of $x$ [Figs.~\ref{figS1}(c)\,--\,(e)].

In addition, there is usually a small NiO phase fraction remaining.

Notably, for these batches, there are no indications of the ``1313'' polymorph of \LNO\ consisting of alternating mono- and trilayers as discussed in the literature \cite{hwang2024a,xchen2024a,puphal2024a}. 

\subsubsection*{\textbf{Lattice parameters}}

\begin{figure}[t]
\centering
\includegraphics[width=0.9\linewidth,clip]{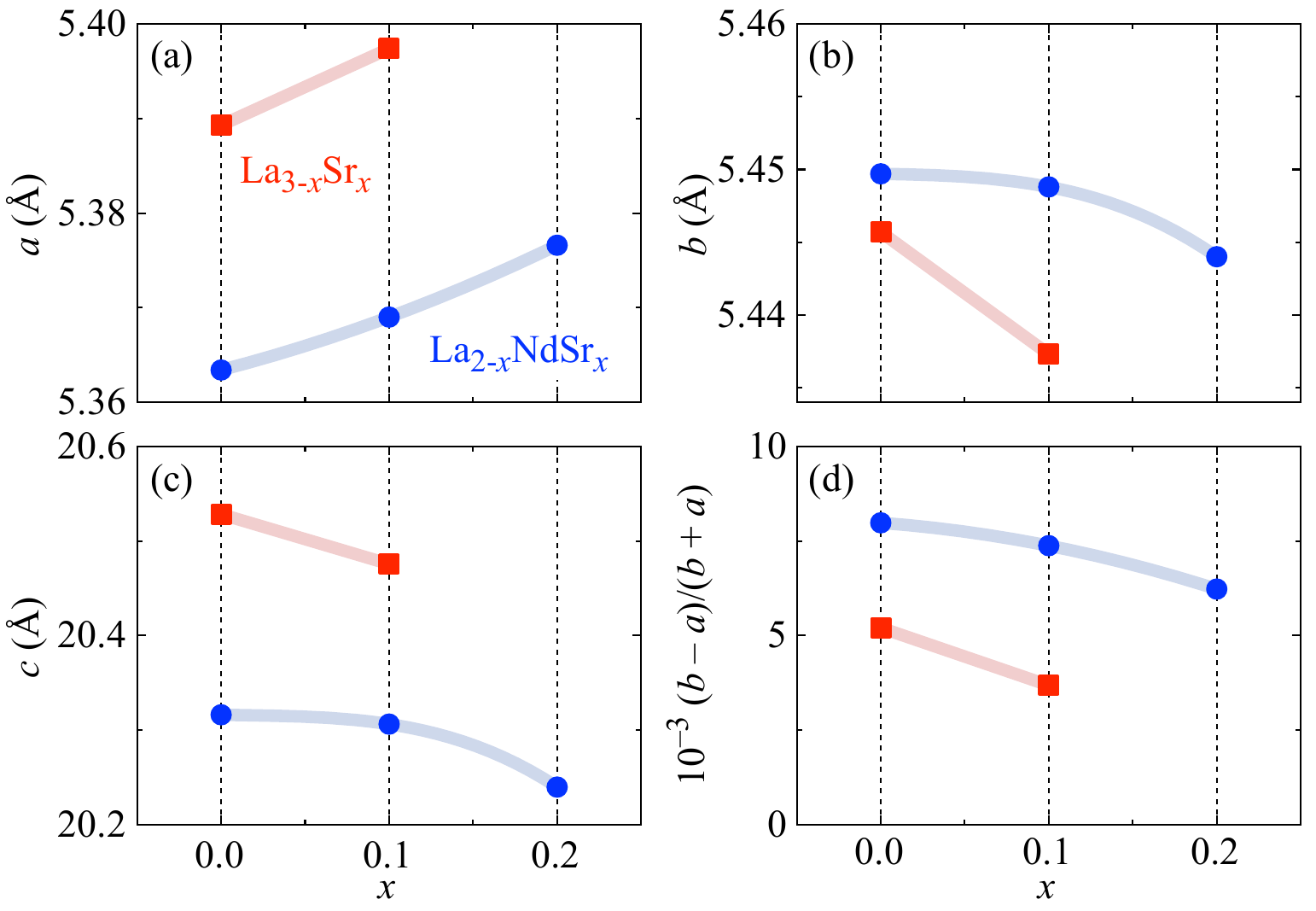}
\caption{(a)\,--\,(c) Lattice parameters $a$, $b$, and $c$, respectively, and (d) orthorhombicity $(b-a)/(b+a)$ as a function of the Sr concentration $x$. Red and blue symbols refer to \LSxNO\ and \LNSxNO, respectively.}
\label{figS2}
\end{figure}

Figures~\ref{figS2}(a)\,--\,(c) summarize the lattice constants $a$, $b$, and $c$, respectively, estimated from the XRD data shown in Fig.~\ref{figS1} and plotted against the Sr concentration $x$. For \LNO, these are in good agreement with the values published in \cite{ling2000a}. When introducing Sr$^{2+}$ with larger ionic radius than La$^{3+}$, the $a$-axis lattice constant is enhanced while $b$ and $c$ shrink with $x$ in \LSxNO. This is in qualitative agreement with the results published on La$_{2.8}$Sr$_{0.2}$Ni$_2$O$_7$ in \cite{mxu2024a}. When introducing smaller Nd$^{3+}$ in \LNNO, the $a$ and $c$ axes become shorter, while $b$ is slightly enhanced. This tendency qualitatively agrees with the case of La$_2$PrNi$_2$O$_7$ in \cite{nwang2024a}, however, the effect on the $b$ axis reported therein is much stronger. 

Doping Sr$^{2+}$ into \LNNO, the same trend as in pristine \LNO\ is observed: The $a$ axis expands, while $b$ and $c$ decrease as a function of $x$, although for $x=0.1$, this effect is more pronounced in \LSxNO\ than in \LNSxNO.

Figure~\ref{figS2}(d) shows the orthorhombicity $(b-a)/(b+a)$ plotted against $x$ which is a measure of the deviation from the higher tetragonal symmetry where $a=b$. This quantity increases upon going from \LNO\ to \LNNO, indicating that the rotation as well as the tilting of the NiO$_6$ octahedra are enhanced when introducing smaller Nd$^{3+}$ ions. Consequently, this trend is reversed as a function of $x$, i.e., doping larger Sr$^{2+}$ ions.

%Figure S2

\clearpage

\subsection{SEM-EDX analyses}
\label{edx}
The chemical compositions of the batches \LRSNO\ discussed in the main text were determined by employing a scanning-electron microscope equipped with an energy-dispersive x-ray analyzer (SEM-EDX; JEOL Ltd.\ and Bruker). 

EDX analyses were performed on different samples from the same batches as used for the transport measurements. On each sample, up to 30 spots in different surface areas were analyzed, and the results averaged with hot spots, e.g., due to NiO impurities excluded. The quantitative results are summarized in Table~\ref{tabS3}. The errors therein are given by the respective standard deviations. Due to the limited sensitivity of EDX to light elements, oxygen is excluded from the quantitative analyses. In this work, the nominal compositions are used when referring to samples. 

\begin{figure}[h!]
\centering
\includegraphics[width=1\linewidth,clip]{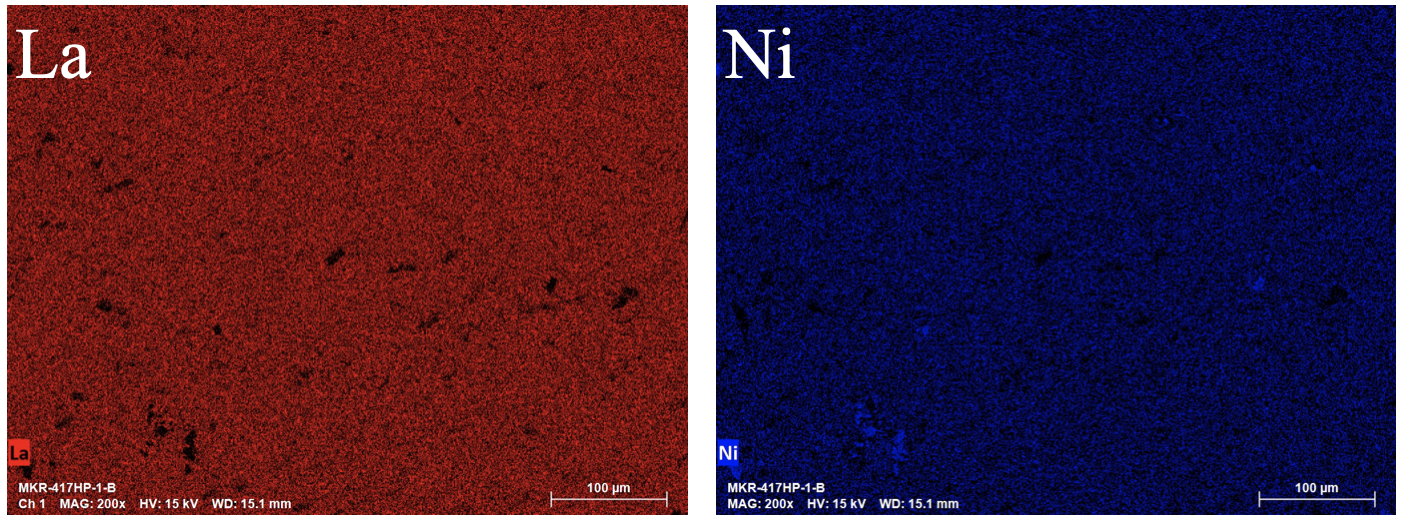}
\caption{Elemental mapping images on \LNO.}
\label{figS3}
\end{figure}

\begin{figure}[h!]
\centering
\includegraphics[width=1\linewidth,clip]{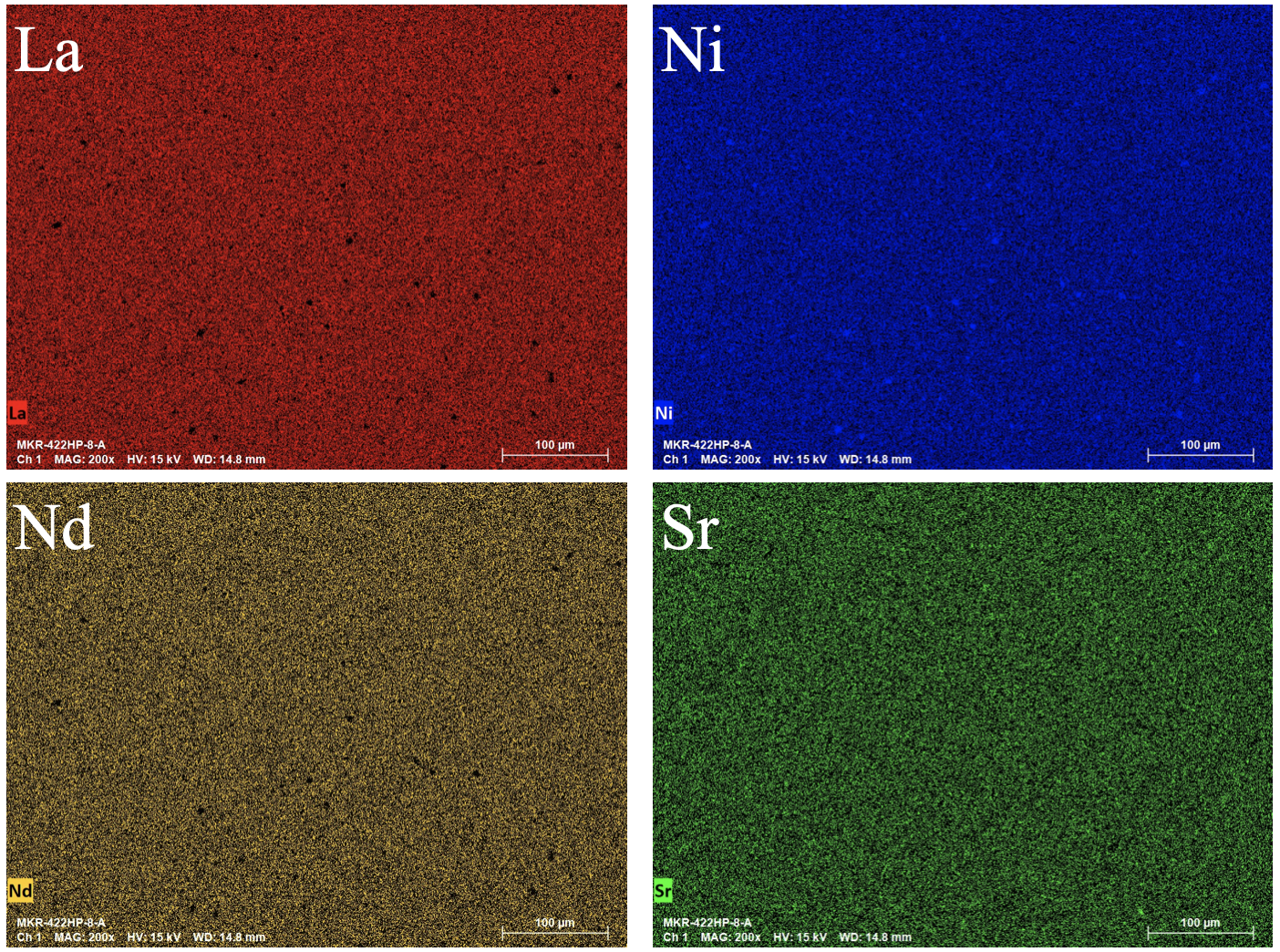}
\caption{Elemental mapping images on \LNSNO. }
\label{figS4}
\end{figure}

In all batches, away from hot spots, the spatial distributions of La, Nd, Sr, and Ni are fairly homogeneous, and the quantitative EDX results are close to the nominal compositions with comparatively small error bars. By contrast, in small but recognizable Ni-rich areas, La and, if included, Nd are depleted, cf.\ Figs.~\ref{figS3} and ~\ref{figS4}. These present representative elemental mapping images for \LNO\ and \LNSNO, respectively.

\begin{table*}[h!]
\centering
\caption{\textbf{Quantitative results of EDX analyses on \texorpdfstring{\boldmath{\LRSNO}}{LRSNO}}}
\label{tabS3}
\setlength{\tabcolsep}{10pt}
\begin{tabular}{ccccc}
\toprule\addlinespace[-0.4em]
 nominal     & \multirow{2}*{La} & \multirow{2}*{Nd} & \multirow{2}*{Sr} & \multirow{2}*{Ni}  \\ \addlinespace[-0.5em]
 composition &                   &                   &                   &                    \\                      \hline \addlinespace[0.5em]
\LNO\        & $2.96 \pm 0.04$   & --                & --                & $2.04 \pm 0.04$    \\ \addlinespace[0.5em] \hline \addlinespace[0.5em]       % MKR-417HP-1
\LSNO\       & $2.90 \pm 0.05$   & --                & $0.09 \pm 0.01$   & $2.01 \pm 0.05$    \\ \addlinespace[0.5em] \hline \addlinespace[0.5em]       % MKR-421HP-2-OX
\LNNO\       & $2.00 \pm 0.04$   & $0.98 \pm 0.03$   & --                & $2.03 \pm 0.07$    \\ \addlinespace[0.5em] \hline \addlinespace[0.3em]       % HM-2-HP-7
\LNSNO\      & $1.86 \pm 0.03$   & $0.97 \pm 0.02$   & $0.12 \pm 0.01$   & $2.03 \pm 0.04$    \\ \addlinespace[0.5em] \hline \addlinespace[0.3em]       % MKR-422HP-8
\LNSSNO\     & $1.78 \pm 0.03$   & $0.98 \pm 0.02$   & $0.20 \pm 0.01$   & $2.03 \pm 0.04$    \\ \addlinespace[0.5em]                                   % MKR-424HP-7
\bottomrule
\end{tabular}
\end{table*}

%Figure S3

%Figure S4

\clearpage

\subsection{Additional STEM results}
\label{stem}

\begin{figure}[b!]
\centering
\includegraphics[width=1\linewidth,clip]{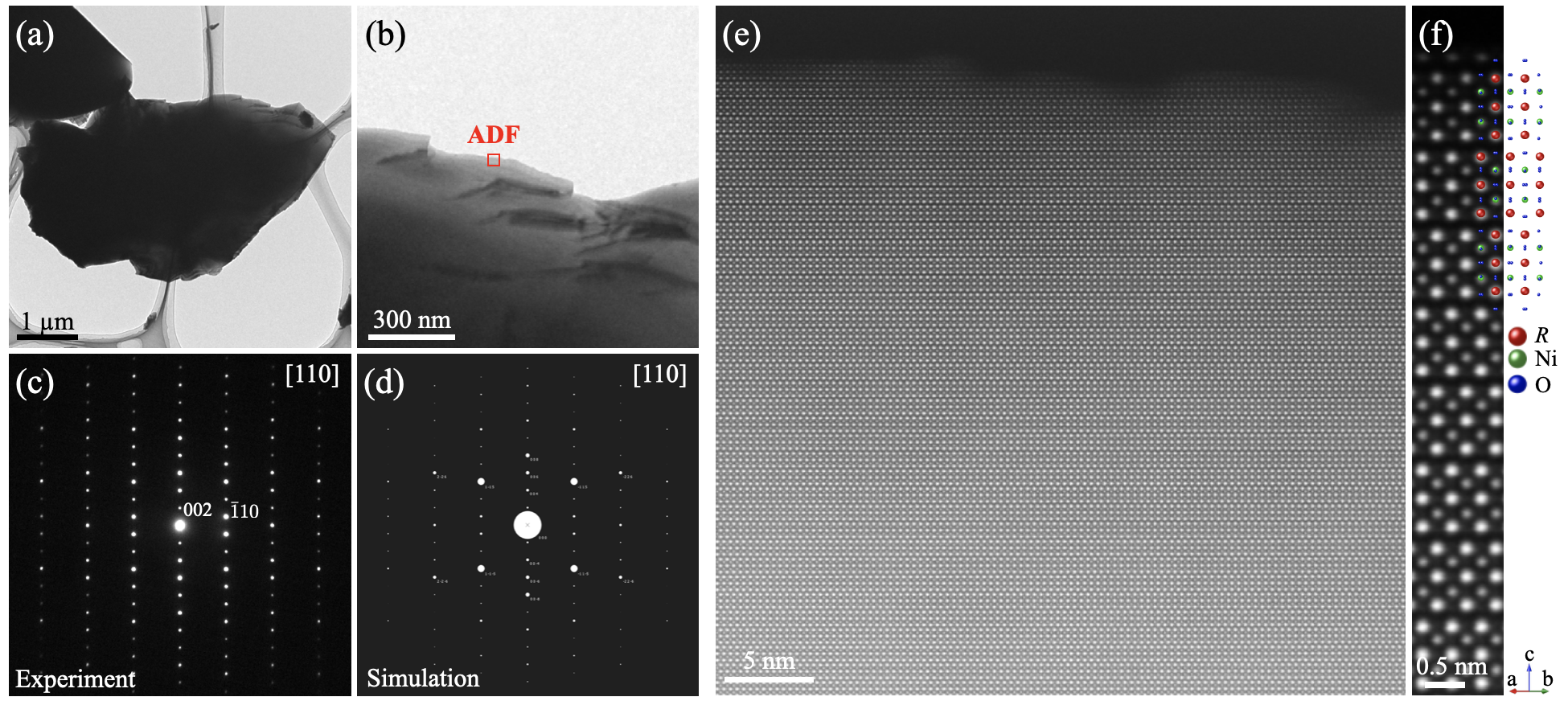}
\caption{Additional scanning transmission electron microscopy (STEM) observations on \LNNO: (a) Bright-field (BF) TEM image of the grain used in the STEM observation. (b) Area on this grain in which annular dark-field (ADF) STEM images were taken. (c) Selected area electron-diffraction pattern obtained on this grain and (d) simulated electron-diffraction pattern viewed along the [110] direction. (e) Atomic-resolution ADF-STEM image viewed along the [110] direction. There are no stacking faults along the $c$ axis discernible. (f) Magnified ADF-STEM image viewed along the [110] direction with the atomic structure model superimposed.}
\label{figS5}
\end{figure}

Figure~\ref{figS5} summarizes additional scanning transmission electron microscopy (STEM) images on \LNNO, as described in the caption. 

A quantitative analysis of the La and Nd occupancy rates on the two lattice sites defined in Fig.~1(a) and visualized in Figs.~1(e) and (f) of the main text yields ratios of
\begin{tabbing}
\hspace{1cm}La\hspace{1cm}         \= : \hspace{0.65em}\= \hspace{0.75cm} Nd\hspace{1.5cm} \= \\
($78.8 \pm 0.8$) \> : \> ($21.2 \pm 0.8$) \> on the $R(1)$ site and \\
($61.0 \pm 0.6$) \> : \> ($39.0 \pm 0.6$) \> on the $R(2)$ site  
\end{tabbing}
for 31 La-rich and 61 La-poor columns, respectively.

%Figure S5

\clearpage

\subsection{Expanded plots of \texorpdfstring{\boldmath{$\rho(T,p)$}}{rho(T,p)}}
\label{rhocomp}

\begin{figure}[b]
\centering
\includegraphics[width=0.8\linewidth,clip]{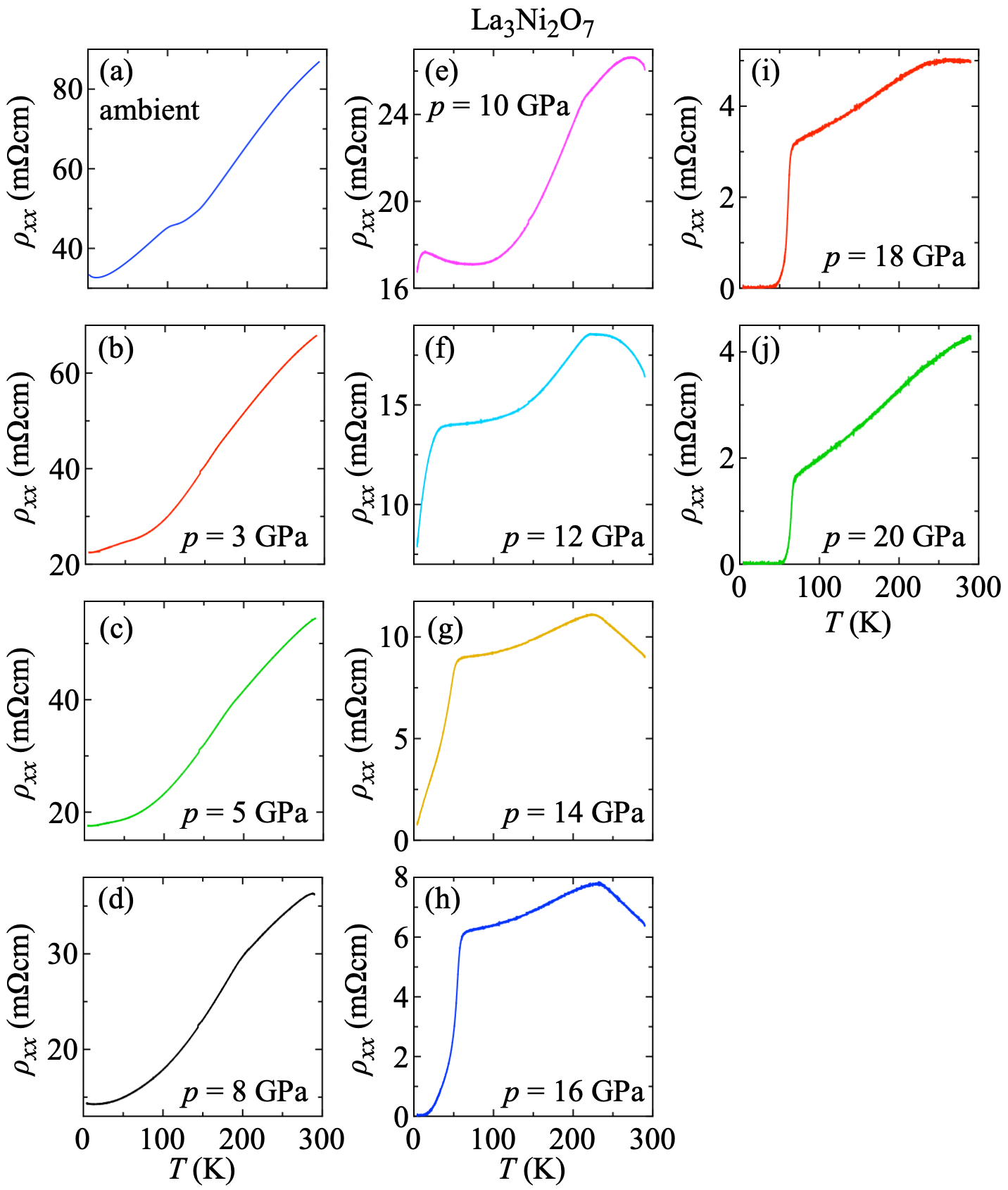}
\caption{Resistivity at ambient and high pressure on \LNO. The same data are shown in Fig.~2(a) of the main text on a logarithmic scale.}
\label{figS6}
\end{figure}

\begin{figure}[t]
\centering
\includegraphics[width=0.7\linewidth,clip]{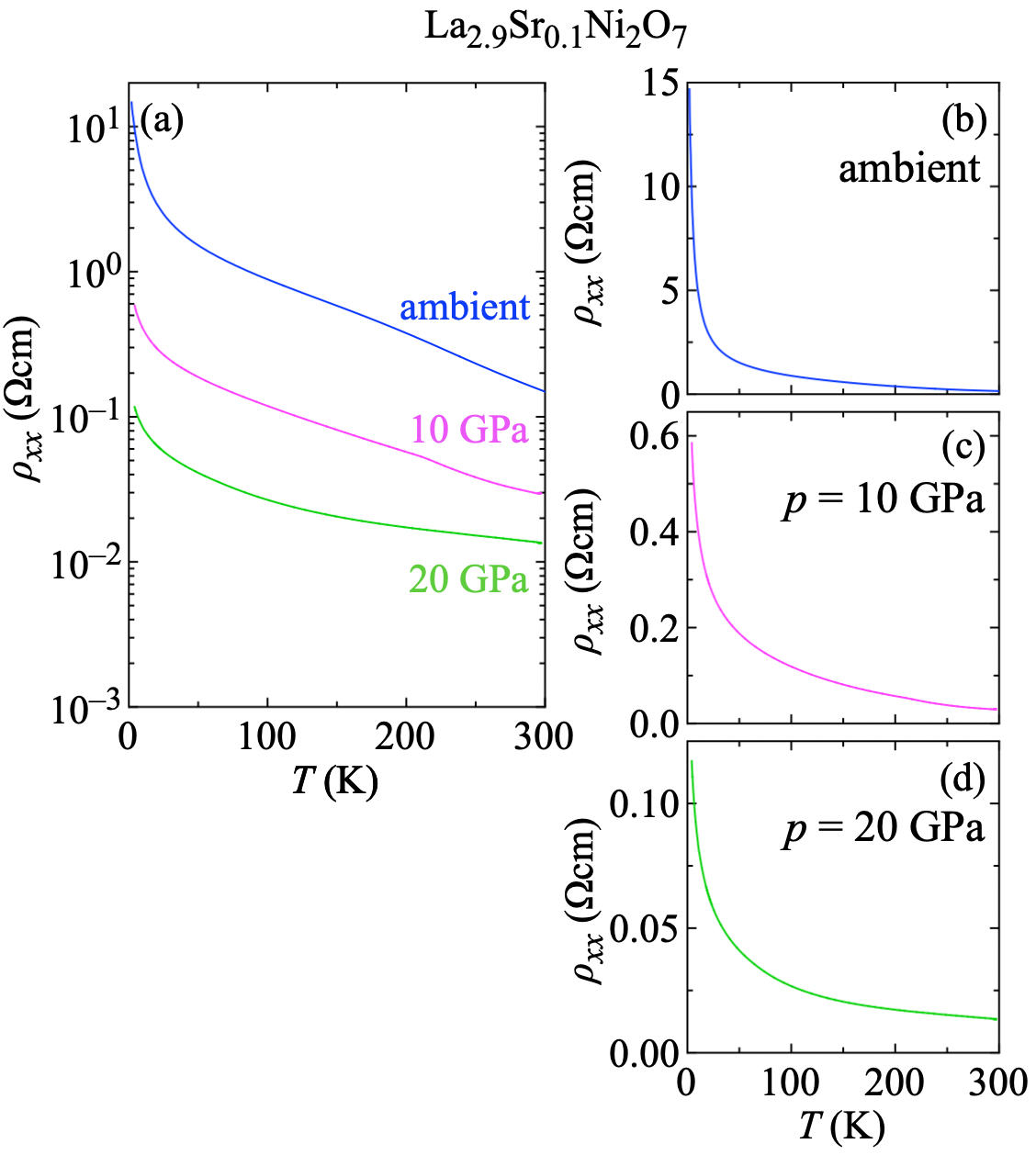}
\caption{(a) Resistivity at ambient and high pressure on \LSNO. (b)\,--\,(d) The same data as in (a) are shown on a linear scale.}
\label{figS7}
\end{figure}

\begin{figure}[t]
\centering
\includegraphics[width=0.6\linewidth,clip]{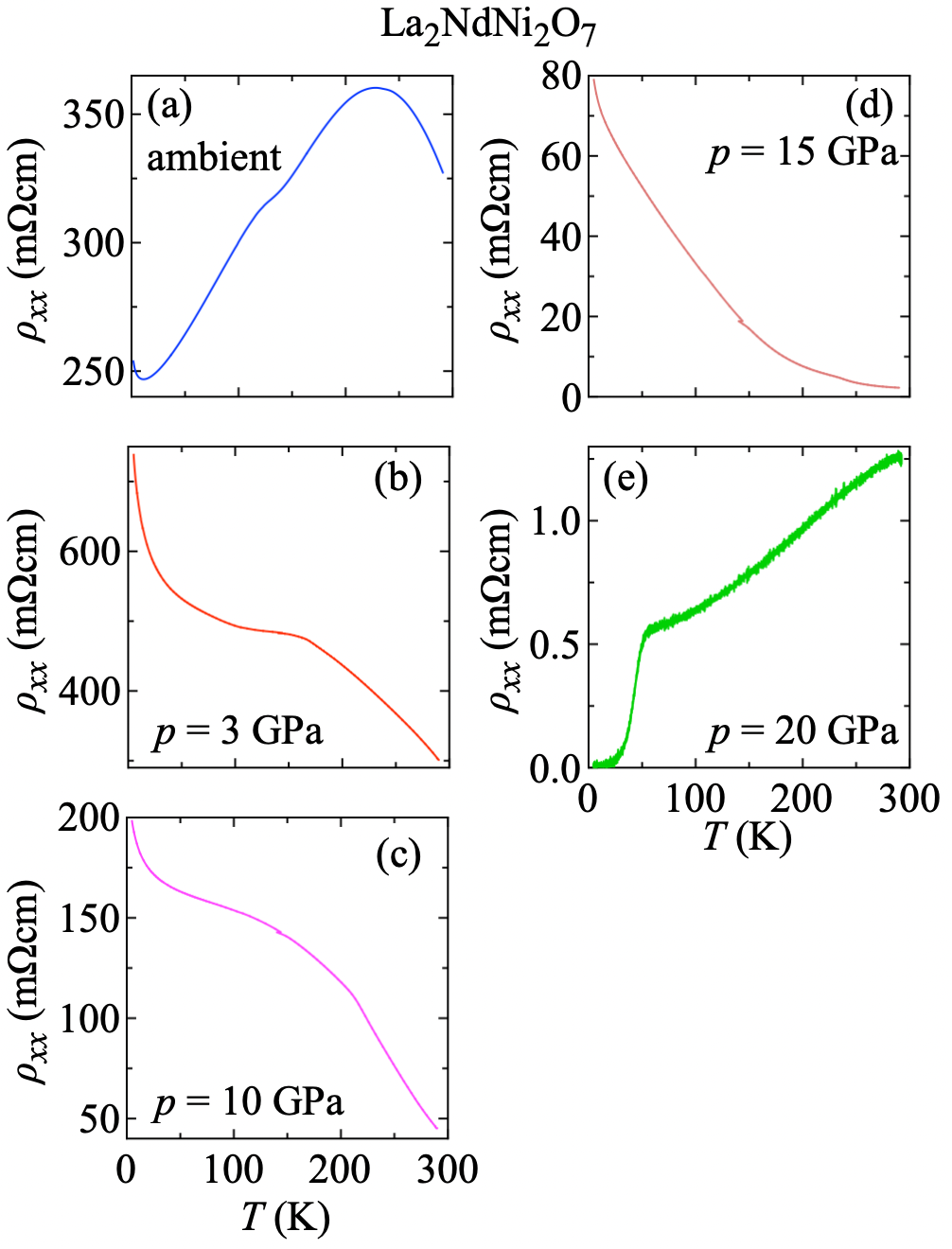}
\caption{Resistivity at ambient and high pressure on \LNNO. The same data are shown in Fig.~2(b) of the main text on a logarithmic scale.}
\label{figS8}
\end{figure}

\begin{figure}[t]
\centering
\includegraphics[width=0.8\linewidth,clip]{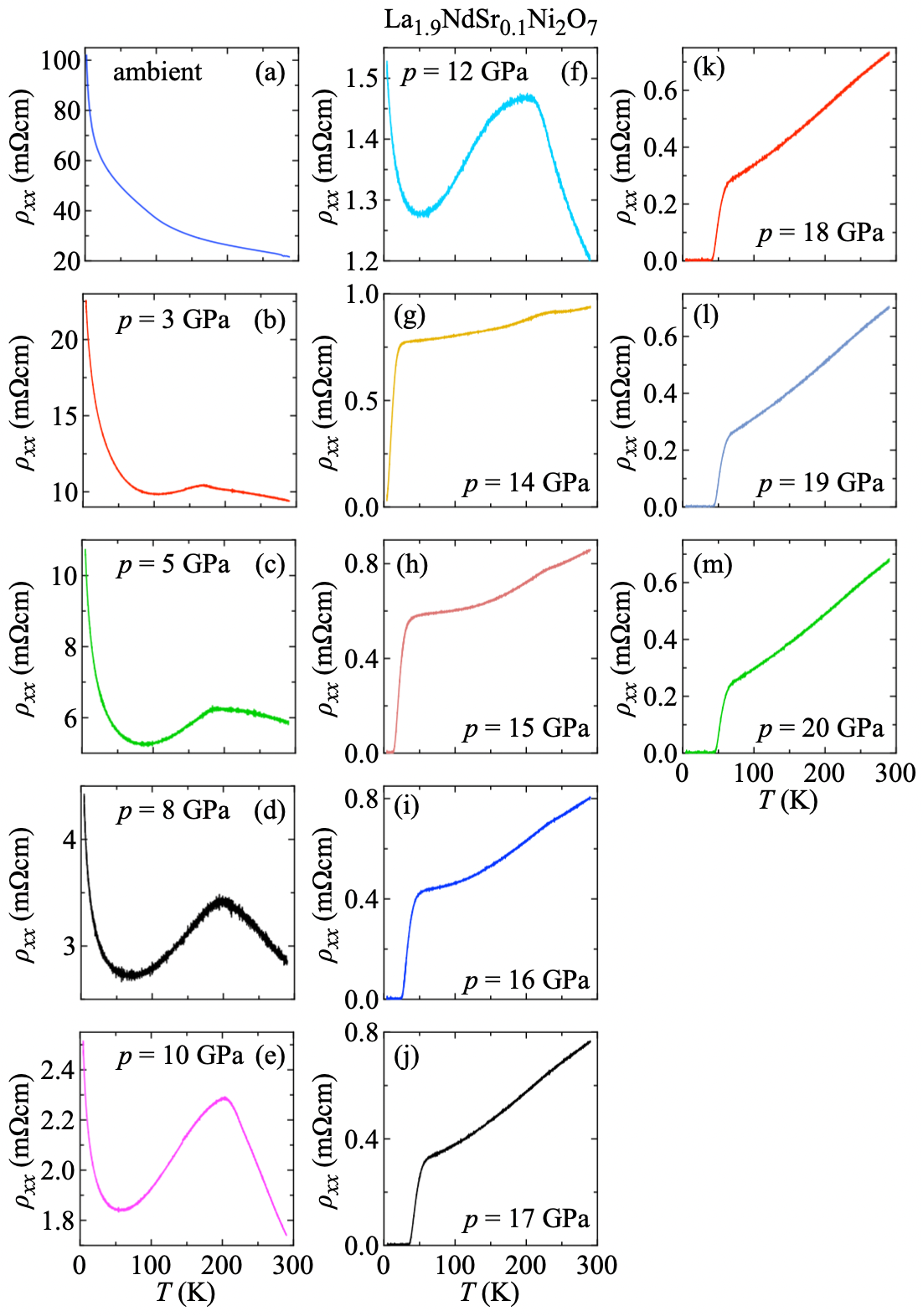}
\caption{Resistivity at ambient and high pressure on \LNSNO. The same data are shown in Fig.~2(c) of the main text on a logarithmic scale except for $p=15$, 17, and 19~GPa, which are omitted therein for clarity.}
\label{figS9}
\end{figure}

\begin{figure}[t]
\centering
\includegraphics[width=0.7\linewidth,clip]{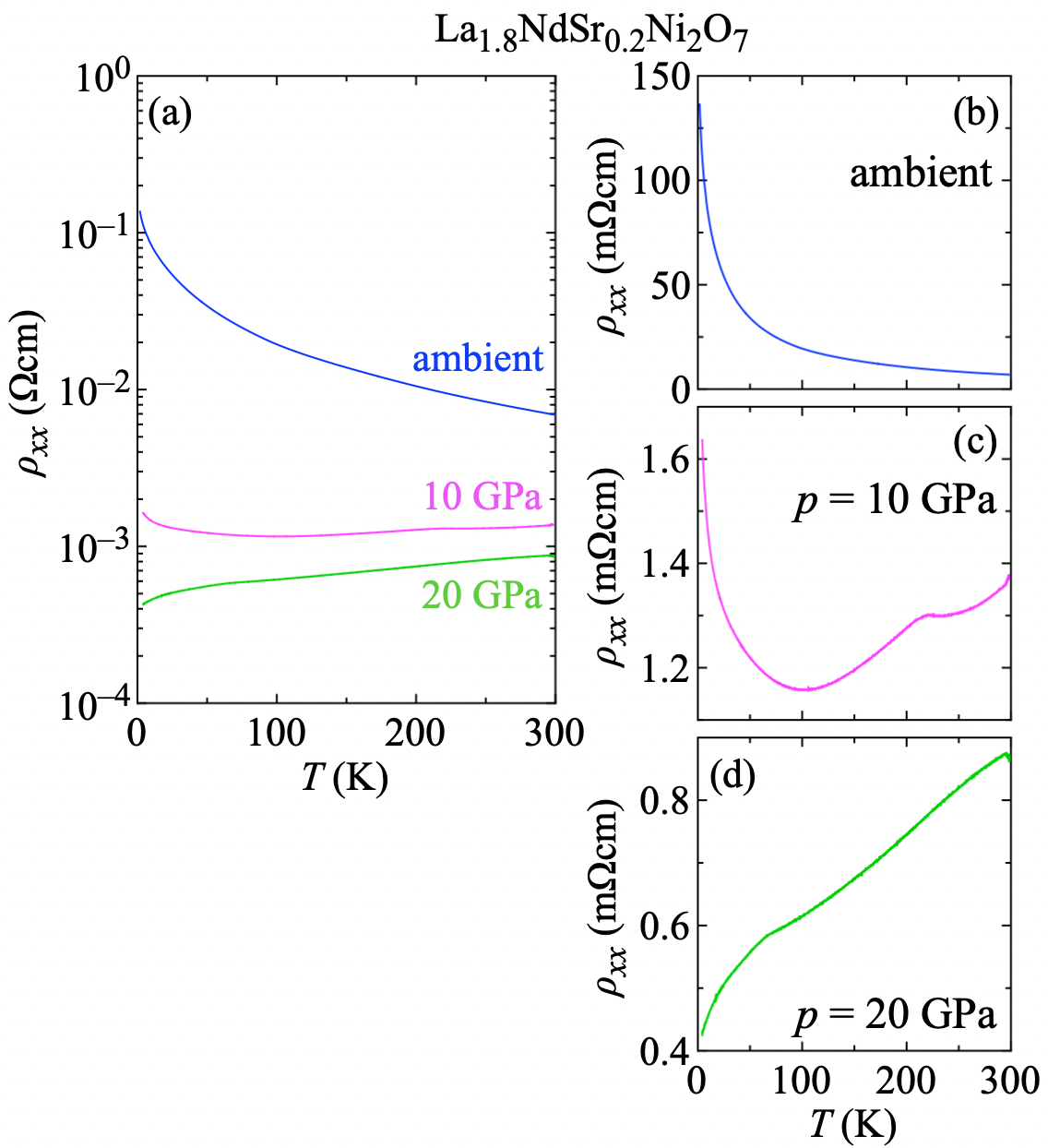}
\caption{(a) Resistivity at ambient and high pressure on \LNSSNO. (b)\,--\,(d) The same data as in (a) are shown on a linear scale. The small downturn below 60~K at 20~GPa in (d) indicates that at this pressure only filamentary superconductivity remains when increasing the Sr concentration to $x=0.2$.}
\label{figS10}
\end{figure}

Figures~\ref{figS6} -- \ref{figS10} provide expanded views on a linear scale of temperature-dependent resistivity data at pressures ranging from ambient to 20~GPa for the five compositions discussed in the main text. This allows a better recognition of the different anomalies and their evolution as functions of $p$ and $T$. For \LSNO\ and \LNSSNO, there are also plots on a logarithmic scale (not shown in the main text). 
%Figure S6

\clearpage

%Figure S7

\clearpage

%Figure S8

\clearpage

%Figure S9

\clearpage

%Figure S10

\clearpage

\subsection{Resistivity \texorpdfstring{\boldmath{$\rho(T)$}}{rho(T)} and temperature derivatives \texorpdfstring{\boldmath{\dr}}{drho/dT} at ambient pressure}
\label{dramb}

%Figure S11
\begin{figure}[h!]
\centering
\includegraphics[width=1\linewidth,clip]{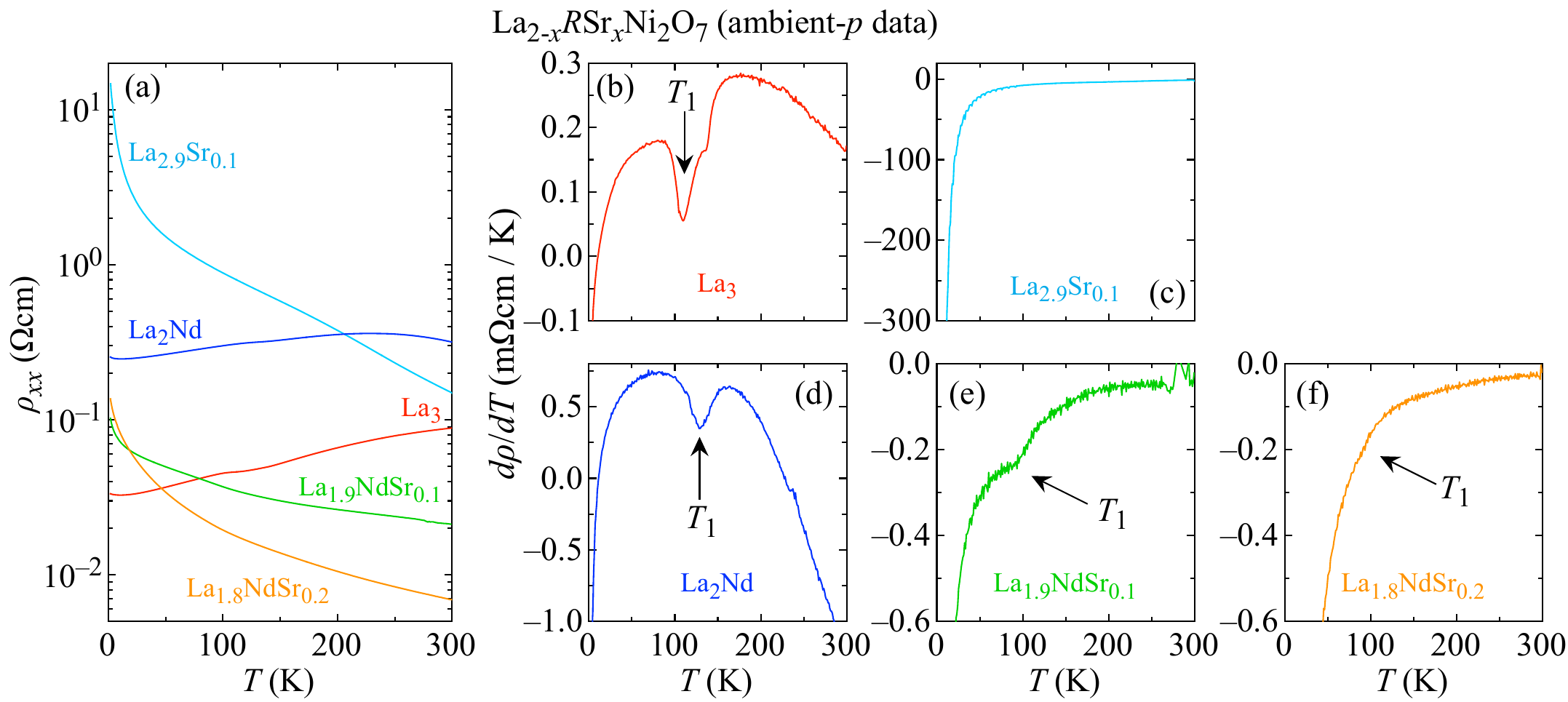}
\caption{(a) Resistivity and (b)\,--\,(f) temperature derivatives \dr\ of the five compositions \LRSNO, as indicated in each panel. All data are taken at ambient pressure. The definitions of the characteristic temperatures $T_1$ are indicated in (b), (d)\,--\,(f). In (c), this type of anomaly is absent. In panels (b)\,--\,(f), \dr\ is displayed in the unit m$\Omega$cm / K as indicated to the left of panels (b) and (d).}
\label{figS11}
\end{figure}

Figure~\ref{figS11}(a) provides a comparison of ambient-pressure \rT\ data of the five compositions \LRSNO\ studied here. The resistivity measured on the \LSNO\ batch (post-growth annealed; Section~\ref{growth}) does not evolve systematically as it exhibits a much larger overall resistivity and a more pronounced localization effect toward low $T$ than observed in \LNSxNO. 

Figures~\ref{figS11}(b)\,--\,(f) present the temperature derivatives \dr\ of the \rT\ data shown in (a). There are strong anomalies $T_1$ in \dr\ for (b) \LNO\ and (d) \LNNO, i.e., the compositions without Sr. In the case of \LNO, introducing Sr suppresses this anomaly, cf.\ panel (c). As for \LNSxNO, the $T_1$ feature is weakened but remains visible for $x=0.1$ [panel (e)], and becomes barely discernible for $x=0.2$ [panel (f)]. This suggests that the introduction of Sr is counteracting the origin of the $T_1$ anomalies, although the data at hand does not allow to conclude on the underlying mechanism. 

Note that in many works on \LNO, the $T_1$ feature at ambient $p$ is allocated at a somewhat higher $T\sim 125 - 140$~K, e.g., \cite{yzhang2024a,khasanov2025a}. This difference stems from the fact that usually \rT\ raw data are used to define $T_1$ which allows some flexibility. The higher temperature range corresponds to a slope change above the flattening in the \rT\ data of \LNO\ shown in Fig.~\ref{figS11}(a), corresponding to the small anomaly seen in panel (b) at $T\sim 137$~K. Since the latter is absent in data obtained on \LNSxNO\ [Figs.~\ref{figS11}(d)\,--\,(f)] and on \LNO\ under pressure (not shown), we define $T_1$ as the most pronounced anomaly in \dr\ for all compositions for consistency.
\clearpage

\subsection{Evolution of \texorpdfstring{\boldmath{$\rho(p)$}}{rho(p)} at \texorpdfstring{\boldmath{$T=10$}}{T = 10}~K for \texorpdfstring{\LNO}{LNO} and \texorpdfstring{\LNSNO}{LNSNO}}
\label{nonmon}

%Figure S12
\begin{figure}[h!]
\centering
\includegraphics[width=0.8\linewidth,clip]{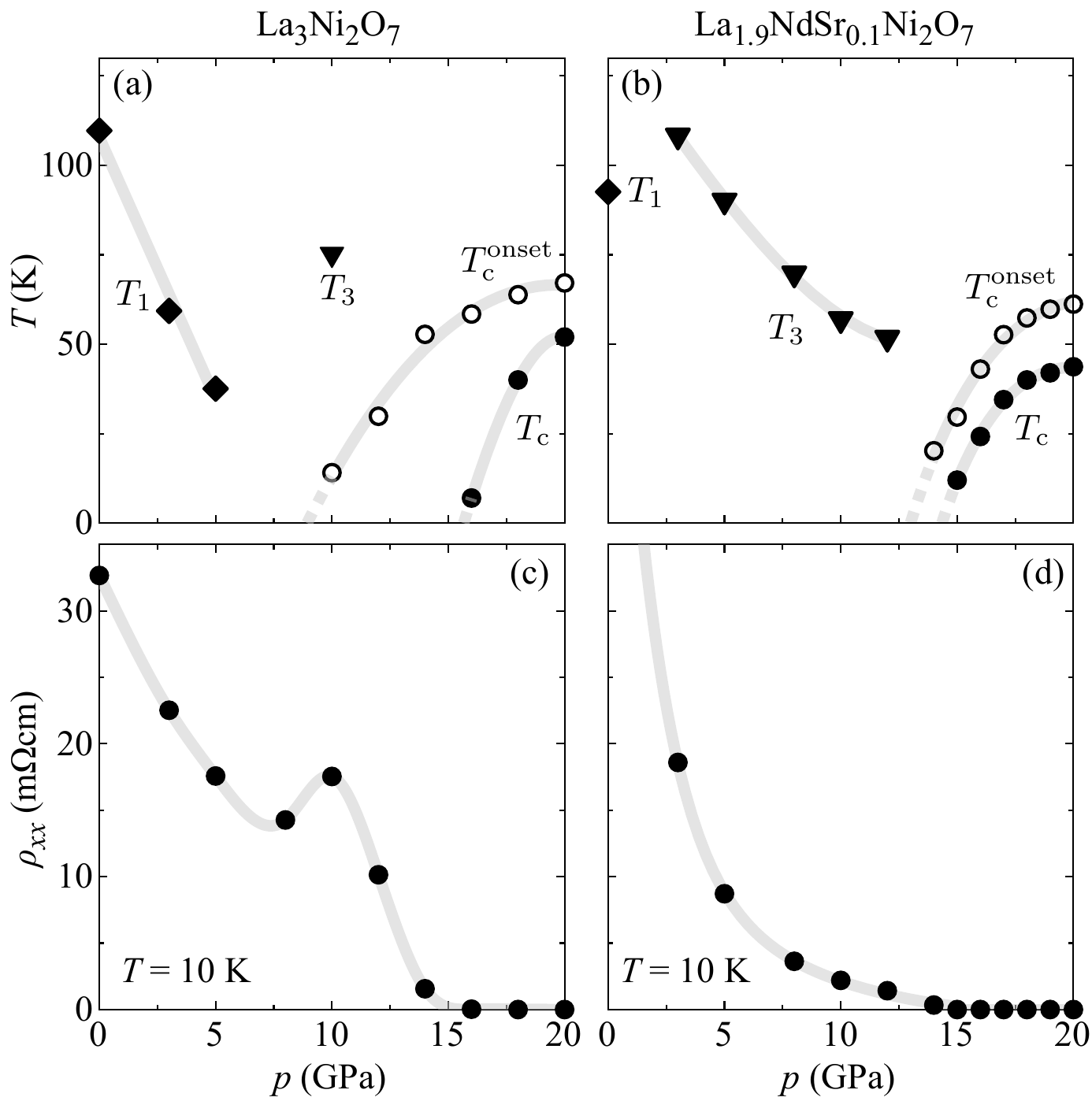}
\caption{(a), (b) Low-temperature region ($T\leq130$~K) of the $\rho(p,T)$ phase diagrams and (c), (d) pressure dependence of the resistivity at $T=10$~K for \LNO\ and \LNSNO, respectively. Thick lines in all panels are guides to the eyes.}
\label{figS12}
\end{figure}

Figures~\ref{figS12}(a) and (b) provide expanded views of the low-temperature region of the \LNO\ and \LNSNO\ phase diagrams shown in Figs.~3(a) and (c) of the main text. The EDX data on \LNSNO\ (Fig.~\ref{figS4}) suggest a homogeneous nature of this batch which appears to be reflected in the nearly pressure-independent transition width between \Ton\ and \Tc\ as seen in Fig.~\ref{figS12}(b).

Fig.~\ref{figS12}(c) summarizes the pressure dependence of the resistivity at $T=10$~K for \LNO, revealing the pronounced nonmonotonic evolution of \rtenK$(p)$ around 10~GPa. This anomaly coincides with the onset of superconductivity in \LNO, cf.\ panel (a), highlighting a change in the electronic system. In Fig.~\ref{figS12}(d), respective data \rtenK$(p)$ are shown for \LNSNO. The systematic and strong suppression of \rtenK\ with $p$ at low $T$ results in a transition from a semiconductor to a metal exhibiting robust superconductivity.

\clearpage